\documentclass{PoS}%
\usepackage{amsmath}
\usepackage{amsfonts}
\usepackage{amssymb}
\usepackage{graphicx}
\usepackage{bm}%
\providecommand{\U}[1]{\protect\rule{.1in}{.1in}}
\title{%
\vspace{-2cm}%
{\small
\begin{flushright} \begin{minipage}{4cm}
KEK preprint 2013-66 \\
CHIBA-EP-205
\end{minipage}\end{flushright}}
\vspace{1cm} %
Non-Abelian dual Meissner effect in SU(3) Yang-Mills theory and confinement/deconfinement phase transition at  finite temperature}

\ShortTitle{Non-Abelian dual Meissner effect in SU(3) Yang-Mills theory and  confinement/deconfinement phase transition}

\author{\speaker{Akihiro Shibata} \\%
Computing Research Center, High Energy Accelerator Research Organization (KEK) \& \\
Graduate University for Advanced Studies (Sokendai), Tsukuba 305-0801, Japan \\
E-mail: \email{akihiro.shibata@kek.jp}
}

\author{Kei-Ichi Kondo \\
Department of Physics, Graduate School of Science, Chiba University, Chiba 263-8522, Japan \\
E-mail: \email{kondok@faculty.chiba-u.jp}
}

\author{Seikou Kato \\
Fukui National College of Technology, Sabae, Fukui 916-8507, Japan \\
E-mail: \email{skato@fukui-nct.ac.jp}
}

\author{Toru Shinohara \\
Department of Physics, Graduate School of Science, Chiba University, Chiba 263-8522, Japan \\
E-mail: \email{sinohara@graduate.chiba-u.jp}
}

\abstract{%
The dual superconductivity is a promising mechanism for quark confinement. 
We have proposed the non-Abelian dual superconductivity picture for SU(3) Yang-Mills theory, 
and showed  the restricted field dominance (called conventionally Abelian dominance), 
and non-Abelian magnetic monopole dominance in the string tension. We have further 
demonstrated by measuring the chromoelectric flux that the non-Abelian dual Meissner effect
exists and determined that the dual superconductivity for SU(3) case is of type I, 
which is in sharp contrast  to the SU(2) case: the border of type I and type II. 

In this talk, we focus on the confinement/deconfinement phase transition  
and the non-Abelian dual superconductivity at a finite temperature: 
We measure the Polyakov loop average and correlator and investigate 
the restricted field dominance in the Polyakov loop. 
Then, we measure the chromoelectric flux between a pair of  static quark and antiquark 
created by a pair of Polyakov loops, and investigate the non-Abelian dual Meissner effect and
its relevance to the phase transition.
}

\FullConference{From quarks and gluons to hadronic matter: A bridge too far ? \\
		 2-6 September, 2013\\
		 European Centre for Theoretical Studies in Nuclear Physics and Related Areas (ECT*), Villazzano, Trento (Italy)}

\begin{document}
\section{Introduction}

The confinement problem is one of the most challenging problems in quantum
chromodynamics (QCD). It has long been argued that the dual superconductivity
is the promising mechanism for quark confinement \cite{dualSC}. In this
scenario, the monopole condensation could play the dominant role for quark
confinement. Quark confinement follows from the area law of the Wilson loop
average, i.e., the string tension for quark and antiquark static potential
must be observed. In many preceding works, the Abelian projection
\cite{tHooft81} was used to show the dual Meissner effect and to perform
numerical analyses, which exhibited the remarkable results: Abelian dominance
\cite{Suzuki90}, magnetic monopole dominance \cite{stack94Shiba}, and center
vortex dominance \cite{greensite} in the string tension. However, these
results are obtained only in special gauges: the maximal Abelian (MA) gauge
and the Laplacian Abelian gauge within the Abelian projection, which breaks
the gauge symmetry as well as color symmetry (global symmetry).

In order to overcome the shortcomings of the Abelian projection and establish
the gauge independent (invariant) mechanism, for quark confinement we have
proposed a new lattice formulation of $SU(N)$ Yang-Mills (YM) theory in the
previous papers \cite{SCGTKKS08L,exactdecomp} (as a lattice version of the
continuum formulations \cite{CFNS-C,KSM05} for $SU(2)$ and
\cite{Cho80c,FN99a,SCGTKKS08} for $SU(N)$), which gives a decomposition of the
gauge link variable suited for extracting the dominant modes for quark
confinement in the gauge independent way. In the case of $SU(2)$, the
decomposition of the gauge link variable was given on a lattice
\cite{KKMSSI06,ref:NLCVsu2,ref:NLCVsu2-2,kato:lattice2009} as a lattice
version of the Cho-Duan-Ge-Faddeev-Niemi-Shabanov (CDGFNS) decomposition
\cite{CFNS-C}. For the gauge group $G=SU(N)$ ($N\geq3$), it was found
\cite{SCGTKKS08} that the extension of the decomposition from $SU(2)$ to
$SU(N)$ ($N\geq3$) is not unique and that there are a number of possible ways
of decompositions discriminated by the stability subgroup $\tilde{H}$ of $G,$
while there is the unique option of $\tilde{H}=U(1)$ in the $SU(2)$ case.

For the case of $G=SU(3)$, in particular, there are two possibilities which we
call the maximal option and the minimal option \cite{SCGTKKS08}. The maximal
option is obtained for the stability group $\tilde{H}=U(1)\times U(1)$, which
enables us to give a gauge invariant version of the MA gauge as the Abelian
projection \cite{lattce2007,Suganuma}. The minimal one is obtained for the
stability group $\tilde{H}=U(2)\cong SU(2)\times U(1)$, which is suited for
representing the Wilson loop in the fundamental representation as derived from
the non-Abelian Stokes theorem \cite{kondo:taira:2000,KondoNAST,KondoShibata}.
In the static potential for a pair of quark and antiquark in the fundamental
representation, we have demonstrated in
\cite{lattice2008,lattice2009,lattice2010} and \cite{abeliandomSU(3)}: (i) the
restricted-field dominance or \textquotedblleft Abelian\textquotedblright%
\ dominance (which is a gauge-independent (invariant) extension of the
conventionally called Abelian dominance): the string tension $\sigma
_{\mathrm{V}}$ obtained from the decomposed $V$-field (i.e., restricted field)
reproduced the string tension $\sigma_{\mathrm{full}}$ of the original YM
field, $\sigma_{\mathrm{V}}/\sigma_{\mathrm{full}}=93\pm16\%$, (ii) the
gauge-independent non-Abelian magnetic monopole dominance: the string tension
$\sigma_{\mathrm{V}}$ extracted from the restricted field was reproduced by
only the (non-Abelian) magnetic monopole part $\sigma_{\mathrm{mon}}$,
$\sigma_{\mathrm{mon}}/\sigma_{\mathrm{V}}=94\pm9\%$.

To establish the non-Abelian dual superconductivity for quark confinement in
$SU(3)$ YM theory which is claimed in \cite{abeliandomSU(3)}, we must
show the evidence of the dual Meissner effect by applying our new formulation
to the $SU(3)$ YM theory on a lattice. In the first half of this talk, we give
a brief review of \cite{DMeisner-TypeI2013}: First, we study the dual Meissner
effect by measuring the distribution of chromo-flux created by a pair of
static quark and antiquark. We compare the chromo-flux of the original
YM field with that of the restricted field and examine if the
restricted field corresponding to the stability group $\tilde{H}=U(2)$
reproduces the dual Meissner effect, namely, the dominant part of the
chromoelectric field strength of $SU(3)$ YM theory. Second, we show
the possible magnetic monopole current induced around the flux connecting a
pair of static quark and antiquark. Third, we focus on the type of dual
superconductivity, i.e., type I or type II. In the $SU(2)$ case, the extracted
field corresponding to the stability group $\tilde{H}=U(1)$ reproduces the
dual Meissner effect, which gives a gauge invariant version of MA gauge in the
Abelian projection. We have shown that the dual superconductivity of the
$SU(3)$ YM theory is indeed the type I, in sharp contrast to the
$SU(2)$ case: the border of type I and type II.

In the latter half of this talk, we study the confinement/deconfinement phase
transition at finite temperature in view of the non-Abelian dual Meissner
effect. We first study our new formulation of lattice YM theory at
finite temperature: by using the decomposed V-field and original YM field, we
measure the space average of Polyakov loops for each configuration, a Polyakov
loop average and correlation functions of the Polyakov loops, and investigate
whether the restricted field (V-field) plays the dominant role at finite
temperature. Then, we measure the distribution of chromo-flux created by a
pair of static quark and antiquark at finite temperature by using both the
restricted field and the original YM field, and examine the chromoelectric
flux tube is generated or not. We find disappearance of the dual Meissner
effect (broken flux tube) at high temperature. Finally, we give summary and outlook

\section{Method}

\subsection{Gauge Link Decomposition}

We introduce a new formulation of the lattice YM theory of the minimal option,
which extracts the dominant mode of the quark confinement for $SU(3)$ YM
theory\cite{abeliandomSU(3),lattice2010}, since we consider the quark
confinement in the fundamental representation. Let $U_{x,\mu}=X_{x,\mu
}V_{x,\mu}$ be the decomposition of YM link variable, where $V_{x.\mu}$ could
be the dominant mode for quark confinement, and $X_{x,\mu}$ the remainder
part. The YM field and the decomposed new-variables are transformed by full
$SU(3)$ gauge transformation $\Omega_{x}$ such that $V_{x,\mu}$ is transformed
as the gauge link variable and $X_{x,\mu}$ as the site available:
\begin{subequations}
\label{eq:gaugeTransf}%
\begin{align}
U_{x,\mu}  &  \longrightarrow U_{x,\nu}^{\prime}=\Omega_{x}U_{x,\mu}%
\Omega_{x+\mu}^{\dag},\\
V_{x,\mu}  &  \longrightarrow V_{x,\nu}^{\prime}=\Omega_{x}V_{x,\mu}%
\Omega_{x+\mu}^{\dag},\text{ \ }X_{x,\mu}\longrightarrow X_{x,\nu}^{\prime
}=\Omega_{x}X_{x,\mu}\Omega_{x}^{\dag}.
\end{align}
The decomposition is given by solving the defining equation:
\end{subequations}
\begin{subequations}
\begin{align}
&  D_{\mu}^{\epsilon}[V]\mathbf{h}_{x}:=\frac{1}{\epsilon}\left[  V_{x,\mu
}\mathbf{h}_{x+\mu}-\mathbf{h}_{x}V_{x,\mu}\right]  =0,\label{eq:def1}\\
&  g_{x}:=e^{i2\pi q/3}\exp(-ia_{x}^{0}\mathbf{h}_{x}-i\sum\nolimits_{j=1}%
^{3}a_{x}^{(j)}\mathbf{u}_{x}^{(i)})=1, \label{eq:def2}%
\end{align}
where $\mathbf{h}_{x}$ is an introduced color field $\mathbf{h}_{x}%
=\xi(\lambda^{8}/2)\xi^{\dag}$ $\in\lbrack SU(3)/U(2)]$ with $\lambda^{8}$
being the Gell-Mann matrix and $\xi$ the $SU(3)$ gauge element. The variable
$g_{x}$ is undetermined parameter from Eq.(\ref{eq:def1}), $\mathbf{u}%
_{x}^{(j)}$ 's are $su(2)$-Lie algebra values, and $q_{x}$ an integer value
$\ 0,1,2$. These defining equations can be solved exactly \cite{exactdecomp},
and the solution is given by
\end{subequations}
\begin{subequations}
\label{eq:decomp}%
\begin{align}
X_{x,\mu}  &  =\widehat{L}_{x,\mu}^{\dag}\det(\widehat{L}_{x,\mu})^{1/3}%
g_{x}^{-1},\text{ \ \ \ }V_{x,\mu}=X_{x,\mu}^{\dag}U_{x,\mu}=g_{x}\widehat
{L}_{x,\mu}U_{x,\mu},\\
\widehat{L}_{x,\mu}  &  =\left(  L_{x,\mu}L_{x,\mu}^{\dag}\right)
^{-1/2}L_{x,\mu},\\
L_{x,\mu}  &  =\frac{5}{3}\mathbf{1}+\frac{2}{\sqrt{3}}(\mathbf{h}%
_{x}+U_{x,\mu}\mathbf{h}_{x+\mu}U_{x,\mu}^{\dag})+8\mathbf{h}_{x}U_{x,\mu
}\mathbf{h}_{x+\mu}U_{x,\mu}^{\dag}\text{ .}%
\end{align}
Note that the above defining equations correspond to the continuum version:
$D_{\mu}[\mathcal{V}]\mathbf{h}(x)=0$ and $\mathrm{tr}(\mathbf{h}%
(x)\mathcal{X}_{\mu}(x))$ $=0,$ respectively. In the naive continuum limit, we
have the corresponding decomposition $\mathbf{A}_{\mathbf{\mu}}(x)=\mathbf{V}%
_{\mu}(x)+\mathbf{X}_{\mu}(x)$ in the continuum theory \cite{SCGTKKS08} as
\end{subequations}
\begin{subequations}
\begin{align}
\mathbf{V}_{\mu}(x)  &  =\mathbf{A}_{\mathbf{\mu}}(x)-\frac{4}{3}\left[
\mathbf{h}(x),\left[  \mathbf{h}(x),\mathbf{A}_{\mathbf{\mu}}(x)\right]
\right]  -ig^{-1}\frac{4}{3}\left[  \partial_{\mu}\mathbf{h}(x),\mathbf{h}%
(x)\right]  ,\\
\mathbf{X}_{\mu}(x)  &  =\frac{4}{3}\left[  \mathbf{h}(x),\left[
\mathbf{h}(x),\mathbf{A}_{\mathbf{\mu}}(x)\right]  \right]  +ig^{-1}\frac
{4}{3}\left[  \partial_{\mu}\mathbf{h}(x),\mathbf{h}(x)\right]  .
\end{align}

The decomposition is uniquely obtained as the solution of Eqs.(\ref{eq:decomp}), 
if color fields$\{\mathbf{h}_{x}\}$ are obtained. To determine the
configuration of color fields, we use the reduction condition which makes the
theory written by new variables ($X_{x,\mu}$,$V_{x,\mu}$) equipollent to the
original YM theory. Here, we use the reduction function
\end{subequations}
\begin{equation}
F_{\text{red}}[\mathbf{h}_{x}]=\sum_{x,\mu}\mathrm{tr}\left\{  (D_{\mu
}^{\epsilon}[U_{x,\mu}]\mathbf{h}_{x})^{\dag}(D_{\mu}^{\epsilon}[U_{x,\mu
}]\mathbf{h}_{x})\right\}  , \label{eq:reduction}%
\end{equation}
and color fields $\left\{  \mathbf{h}_{x}\right\}  $ are obtained by
minimizing the functional.

\subsection{Chromo-flux created by a pair of quark and antiquark}

\begin{figure}[ptb]
\begin{center}
\includegraphics[
height=4.5cm]
{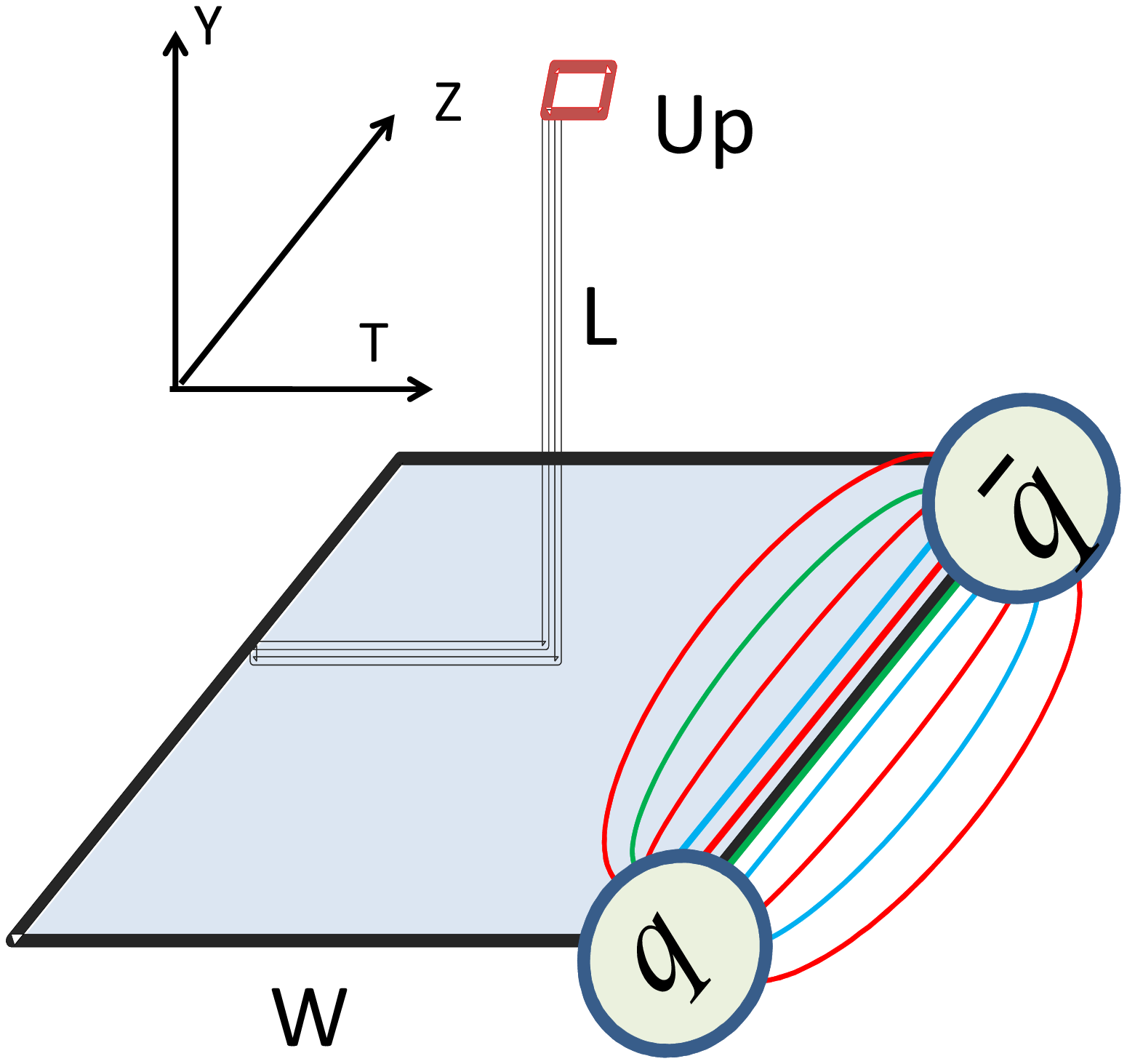} \ \ \ \ \ \ \ \includegraphics[
height=3.8cm]
{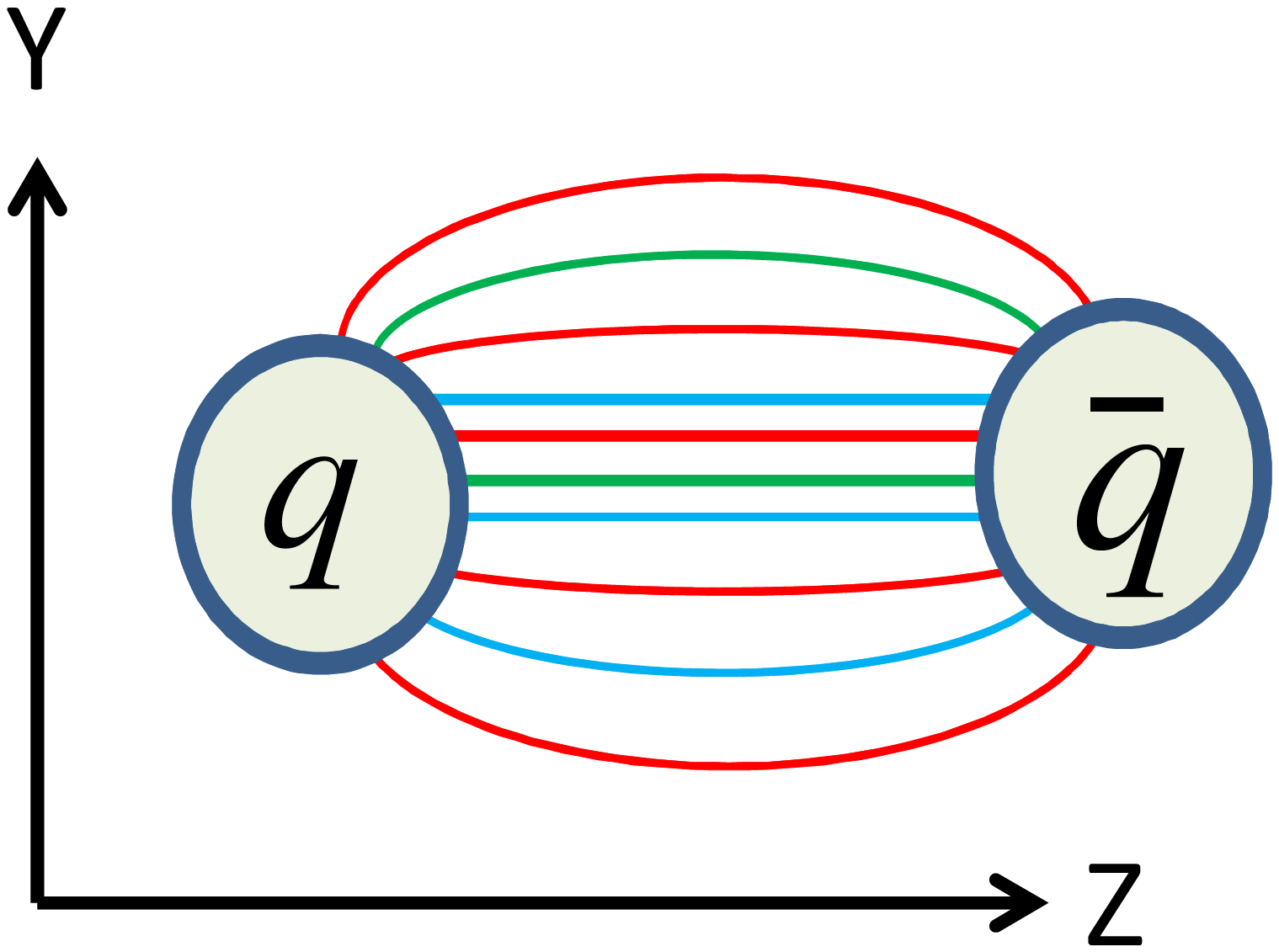}
\end{center}
\caption{(Left) The connected correlator ($U_{p}LWL^{\dag})$ between a
plaquette $U_{p}$ and the Wilson loop $W$. (Right) Measurement of  the chromo
flux in the Y-Z plane. }%
\label{fig:Operator}%
\end{figure}

We investigate the non-Abelian dual Meissner effect as the mechanism of quark
confinement. In order to extract the chromo-field, we use a gauge-invariant
correlation function proposed in \cite{Giacomo}: The gauge-invariant
chromo-field strength $F_{\mu\nu}[U]$ created by a quark-antiquark pair in
$SU(3)$ YM theory is measured by using a gauge-invariant connected correlator
between a plaquette and the Wilson loop (see Fig.\ref{fig:Operator}):%
\begin{equation}
F_{\mu\nu}[U]:=\epsilon^{-2}\sqrt{\frac{\beta}{6}}\rho_{W}[U],\text{
\ \ \ \ \ \ \ \ }\rho_{W}[U]:=\frac{\left\langle \mathrm{tr}\left(
U_{P}L[U]^{\dag}W[U]L[U]\right)  \right\rangle }{\left\langle \mathrm{tr}%
\left(  W[U]\right)  \right\rangle }-\frac{1}{3}\frac{\left\langle
\mathrm{tr}\left(  U_{P}\right)  \mathrm{tr}\left(  W[U]\right)  \right\rangle
}{\left\langle \mathrm{tr}\left(  W[U]\right)  \right\rangle },\label{eq:Op}%
\end{equation}
where $\beta:=6/g^{2}$ is the lattice gauge coupling constant, $W$ the Wilson
loop in $Z$-$T$ plane representing a pair of quark and antiquark, $U_{P}$ a
plaquette variable as the probe operator to measure the chromo-field strength
at the point $P$, and $L$ the Wilson line connecting the source $W$ and the
probe $U_{P}$. Here $L$ is necessary to guarantee the gauge invariance of the
correlator $\rho_{W}$ and hence the probe is identified with $LU_{P}%
L^{\dagger}$.
The symbol $\left\langle \mathcal{O}\right\rangle $ denotes the average of the
operator $\mathcal{O}$ in the space and the ensemble of the configurations. In
the naive continuum limit $\epsilon\rightarrow0$, indeed, $\rho_{W}$ reduces
to the field strength in the presence of the $q\bar{q}$ source: $\rho
_{W}\overset{\varepsilon\rightarrow0}{\simeq}g\epsilon^{2}\left\langle
\mathcal{F}_{\mu\nu}\right\rangle _{q\bar{q}}:=\frac{\left\langle
\mathrm{tr}\left(  ig\epsilon^{2}L\mathcal{F}_{\mu\nu}L^{\dag}W\right)
\right\rangle }{\left\langle \mathrm{tr}\left(  W\right)  \right\rangle
}+O(\epsilon^{4}),$where we have used $U_{x,\mu}=\exp(-ig\epsilon
\mathcal{A}_{\mu}(x))$ and hence $U_{P}=\exp(-ig\epsilon^{2}\mathcal{F}%
_{\mu\nu})$. 

\begin{figure}[ptb]
\begin{center}
\includegraphics[
width=4cm,
angle=270,
]
{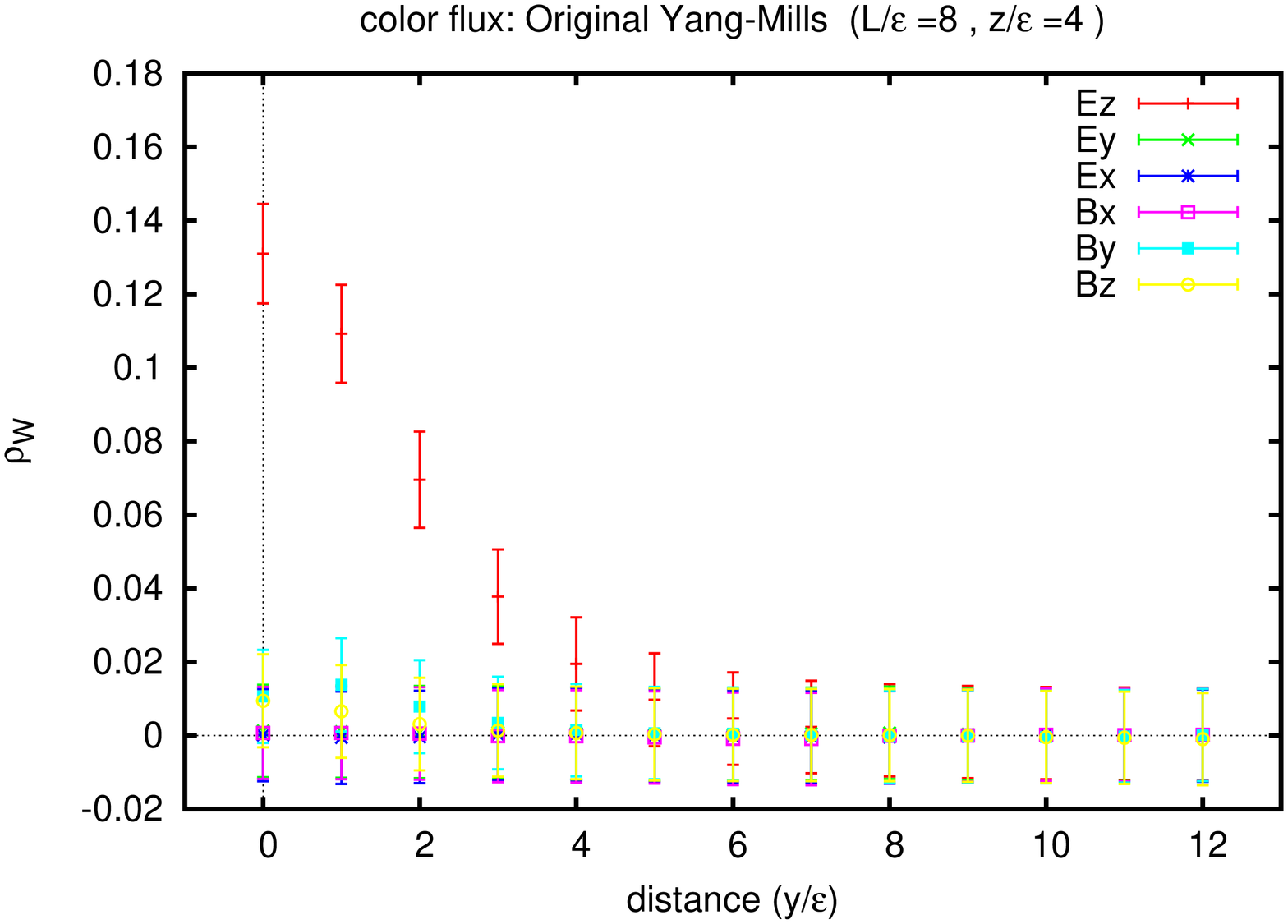} \ \ \includegraphics[
width=4cm,
angle=270,
]
{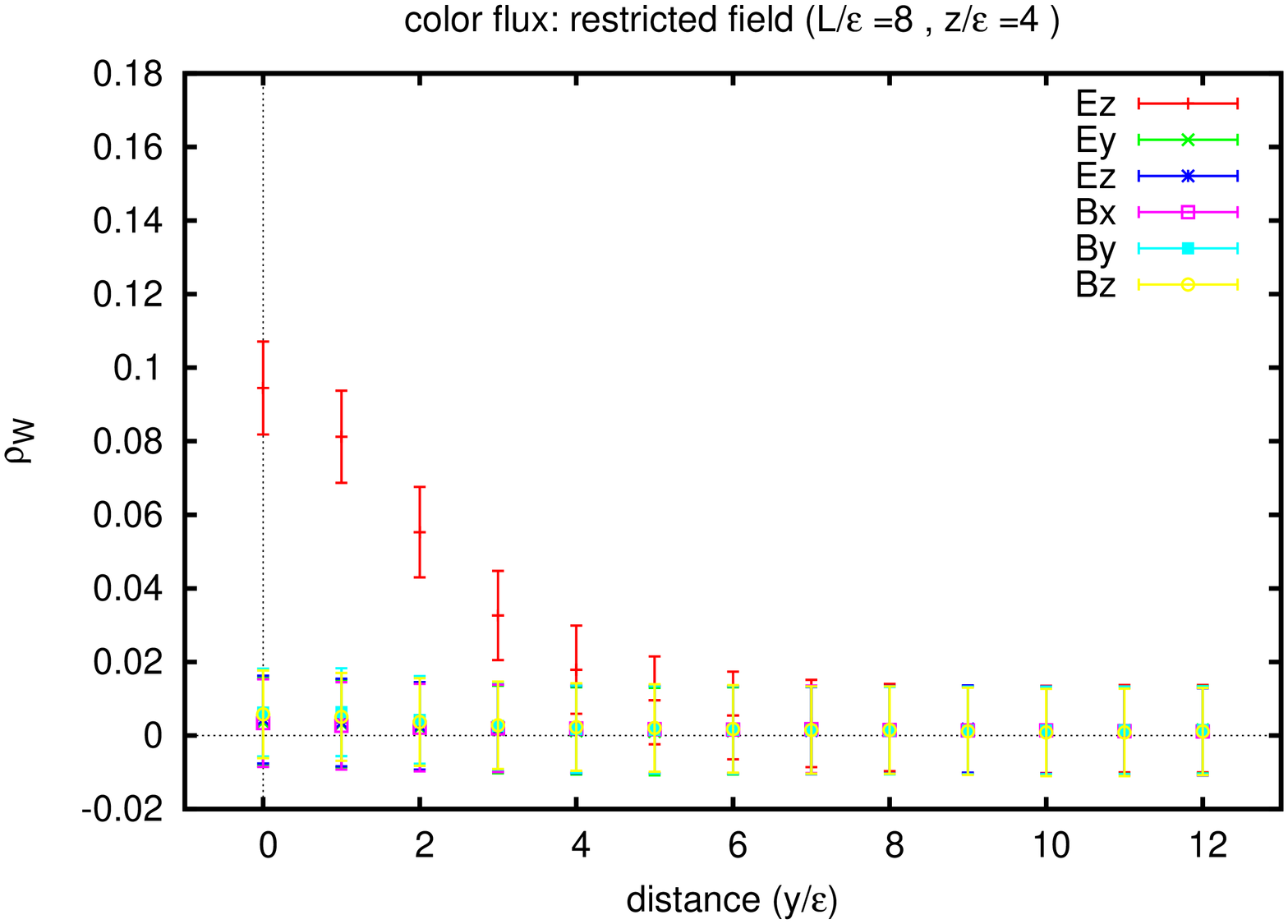}
\end{center}
\caption{ Measurement of components of the chromoelectric field $\bm{E}$ and
chromomagnetic field $\bm{B}$ as functions of the distance $y$ from the $z$
axis. (Left panel) the original $SU(3)$ YM field, (Right panel) the restricted
field. }%
\label{fig:measure}%
\end{figure}

We measure correlators between the plaquette $U_{P}$ and the chromo-field
strength of the restricted field $V_{x,\mu}$ as well as the original YM field
$U_{x,\mu}$. See the left panel of Fig.~\ref{fig:Operator}. Here the quark and
antiquark source is introduced as $8\times8$ Wilson loop ($W$) in the $Z$-$T$
plane, and the probe $(U_{p})$ is set at the center of the Wilson loop and
moved along the $Y$-direction. The left and right panel of
Fig.~\ref{fig:measure} show respectively the results of measurements for the
chromoelectric and chromomagnetic fields $F_{\mu\nu}[U]$ for the original
$SU(3)$ field $U$ and $F_{\mu\nu}[V]$ for the restricted field $V$, where the
field strength $F_{\mu\nu}[V]$ is obtained by using $V_{\,x,\mu}$ in
eq(\ref{eq:Op}) instead of $U_{x,\mu}$:
\begin{equation}
F_{\mu\nu}[V]:=\sqrt{\frac{\beta}{6}}\rho_{W}[V],\quad\rho_{W}[V]:=\frac
{\left\langle \mathrm{tr}\left(  L[V]V_{p}L^{\dag}[V]W[V]\right)
\right\rangle }{\left\langle \mathrm{tr}\left(  W[V]\right)  \right\rangle
}-\frac{1}{3}\frac{\left\langle \mathrm{tr}\left(  V_{p}\right)
\mathrm{tr}\left(  W[V]\right)  \right\rangle }{\left\langle \mathrm{tr}%
\left(  W[V]\right)  \right\rangle }.
\end{equation}
We have checked that even if $W[U]$ is replaced by $W[V]$, together with
replacement of the probe $LU_{P}L^{\dagger}$ by the corresponding $V$ version,
the change in the magnitude of the field strength $F_{\mu\nu}$ remains within
at most a few percent.

\section{Measurement of chromo flux on a lattice}

We generate configurations of the YM gauge link variable $\{U_{x,\mu}\}$ using
the standard Wilson gauge action on $L^{3}\times N_{T}$ lattice at $\beta$:
$L=24$, $N_{T}=6,8,10,14,24$ with $\beta=6.0$; $L=24$,$\ N_{T}%
=4,6,8,10,12,14,24$ with $\beta=6$.$2$, and $L=24$, $N_{T}=4,6$ with
$\beta=6.4$. The gauge link decomposition $U_{x,\mu}=X_{x,\mu}V_{x,\mu}$ is
obtained by the formula given in the previous section: the color field
configuration $\{h_{x}\}$ is obtained by solving the reduction condition of
minimizing the functional eq.(\ref{eq:reduction}) for each gauge configuration
$\{U_{x,\mu}\}$, and then the decomposed variables $\{V_{x,\mu}\}$,
$\{X_{x,\mu}\}$ are obtained by using the formula eq.(\ref{eq:decomp}). In the
measurement of the Polyakov loop and Wilson loop, we apply the APE smearing
technique to reduce noises \cite{APEsmear}.

\subsection{Non-Abelian dual Meissner effect at zero temperature}

\subsubsection{Chromo flux tube}

From Fig.\ref{fig:measure} we find that only the $E_{z}$ component of the
chromoelectric field $(E_{x},E_{y},E_{z})=(F_{10},F_{20},F_{30})$ connecting
$q$ and $\bar{q}$ has non-zero value for both the restricted field $V$ and the
original YM field $U$. The other components are zero consistently within the
numerical errors. This means that the chromomagnetic field $(B_{x},B_{y}%
,B_{z})=(F_{23},F_{31},F_{12})$ connecting $q$ and $\bar{q}$ does not exist
and that the chromoelectric field is parallel to the $z$ axis on which quark
and antiquark are located. The magnitude $E_{z}$ quickly decreases in the
distance $y$ away from the Wilson loop.

\begin{figure}[ptb]
\begin{center}
\includegraphics[
height=5.0cm,
angle=270
]
{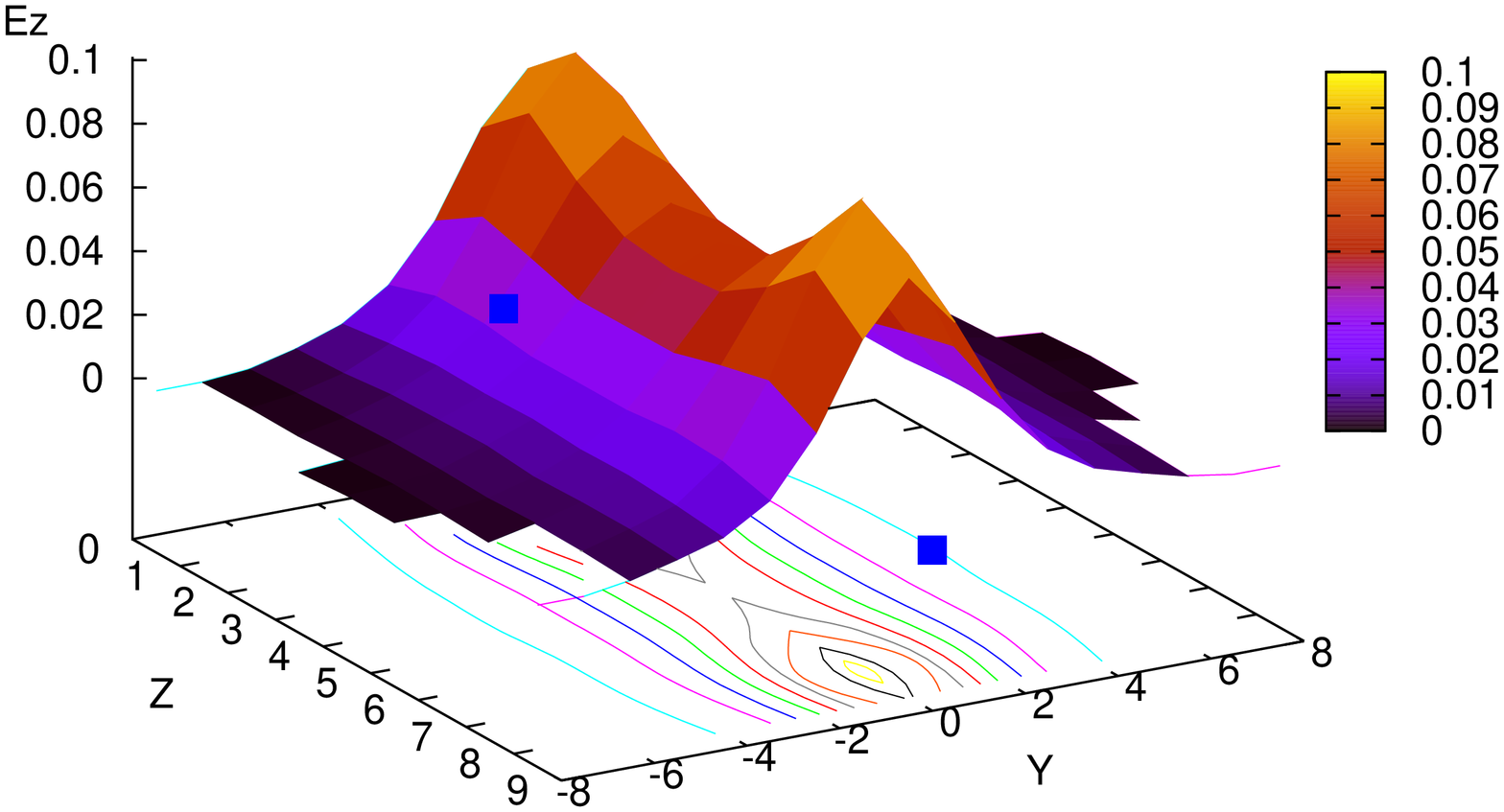} \quad\includegraphics[
height=5.0cm,
angle=270
]
{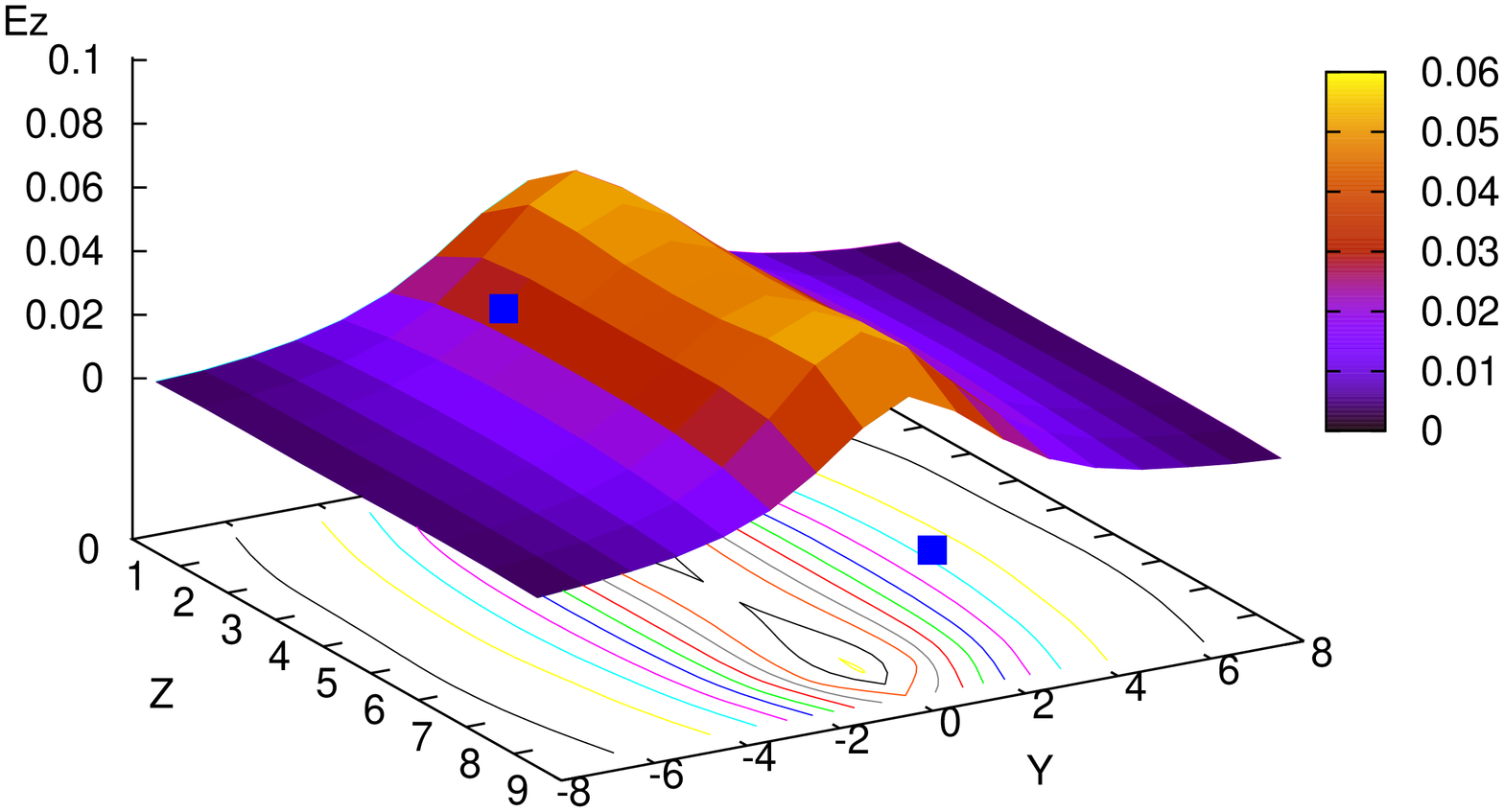} \vspace{-5mm}
\end{center}
\caption{ The distribution in $Y$-$Z$ plane of the chromoelectric field
$E_{z}$ connecting a pair of quark and antiquark: (Left panel) chromoelectric
field produced from the original YM field, (Right panel) chromoelectric field
produced from the restricted $U(2)$ field. }%
\label{fig:fluxtube}%
\end{figure}

To see the profile of the nonvanishing component $E_{z}$ of the chromoelectric
field in detail, we explore the distribution of chromoelectric field on the
2-dimensional plane. Fig.~\ref{fig:fluxtube} shows the distribution of $E_{z}$
component of the chromoelectric field, where the quark-antiquark source
represented as $9\times11$ Wilson loop $W$ is placed at $(Y,Z)=(0,0),(0,9)$,
and the probe $U$ is displaced on the $Y$-$Z$ plane at the midpoint of the
$T$-direction. The position of a quark and an antiquark is marked by the solid
(blue) box. The magnitude of $E_{z}$ is shown by the height of the 3D plot and
also the contour plot in the bottom plane. The left panel of
Fig.~\ref{fig:fluxtube} shows the plot of $E_{z}$ for the $SU(3)$ YM field
$U$, and the right panel of Fig.~\ref{fig:fluxtube} for the restricted-field
$V$. We find that the magnitude $E_{z}$ is quite uniform for the restricted
part $V$, while it is almost uniform for the original part $U$ except for the
neighborhoods of the locations of $q$, $\bar{q}$ source. This difference is
due to the contributions from the remaining part $X$ which affects only the
short distance.\cite{DMeisner-TypeI2013}

\subsubsection{Magnetic current}

Next, we investigate the relation between the chromoelectric flux and the
magnetic current. The magnetic(-monopole) current can be calculated as
\begin{equation}
\mathbf{k}={}^{\ast}dF[\mathbf{V}], \label{def-k}%
\end{equation}
where $F[\mathbf{V}]$ is the field strength (2-form) of the restricted field
(1-form) $\mathbf{V}$, $d$ the exterior derivative and $^{\ast}$ denotes the
Hodge dual operation. Note that non-zero magnetic current follows from
violation of the Bianchi identity (If the field strength was given by the
exterior derivative of $\mathbf{V}$ field (one-form), $F[\mathbf{V}%
]=d\mathbf{V}$, \ we would obtain $\mathbf{k=}^{\ast}d^{2}\mathbf{V}$ $=0$).

\begin{figure}[ptb]
\begin{center}
\includegraphics[
width=4.5cm
]
{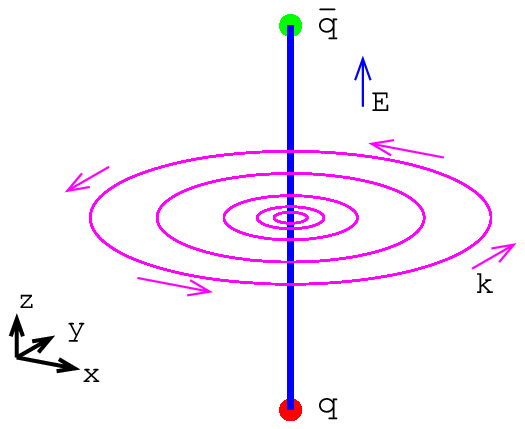} \quad\includegraphics[
width=6cm
]
{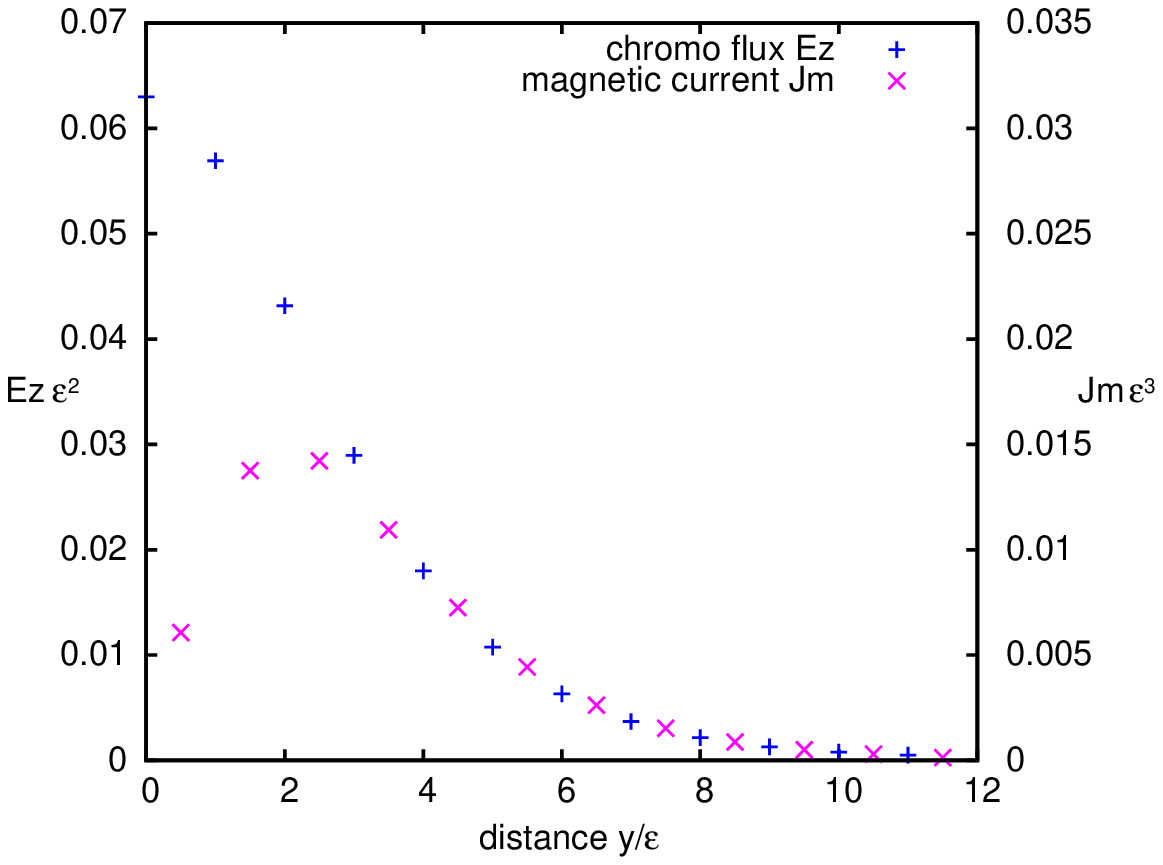} \vspace{-5mm}
\end{center}
\caption{{}The magnetic-monopole current $\mathbf{k}$ induced around the flux
along the $z$ axis connecting a quark-antiquark pair. (Left panel) The
positional relationship between the chromoelectric field $E_{z}$ and the
magnetic current $\mathbf{k}$. (Right panel) The magnitude of the
chromo-electronic current $E_{z}$ and the magnetic current $J_{m}%
=|\mathbf{k}|$ as functions of the distance $y$ from the $z$ axis. }%
\label{fig:Mcurrent}%
\end{figure}

Fig.~\ref{fig:Mcurrent} shows the magnetic current measured in $X$-$Y$ plane
at the midpoint of quark and antiquark pair in the $Z$-direction. The left
panel of Fig.~\ref{fig:Mcurrent} shows the positional relationship between
chromoelectric flux and magnetic current. The right panel of
Fig.~\ref{fig:Mcurrent} shows the magnitude of the chromoelectric field
$E_{z}$ (left scale) and the magnetic current $k$ (right scale). The existence
of nonvanishing magnetic current $k$ around the chromoelectric field $E_{z}$
supports the dual picture of the ordinary superconductor exhibiting the
electric current $J$ around the magnetic field $B$.

In our formulation, it is possible to define a gauge-invariant
magnetic-monopole current $k_{\mu}$ by using $V$-field,
which is obtained from the field strength $\mathcal{F}_{\mu\nu}[\mathbf{V}]$
of the field $\mathbf{V}$, as suggested from the non-Abelian Stokes theorem
\cite{KondoNAST,KondoShibata}. It should be also noticed that this
magnetic-monopole current is a non-Abelian magnetic monopole extracted from
the $V$ field, which corresponds to the stability group $\tilde{H}=U(2)$. The
magnetic-monopole current $k_{\mu}$ defined in this way can be used to study
the magnetic current around the chromoelectric flux tube, instead of the above
definition of $k$,  Eq.(\ref{def-k}).

\subsubsection{Type of dual superconductivity}

\begin{figure}[ptb]
\begin{center}
\includegraphics[
width=5.0cm,
angle=270
]
{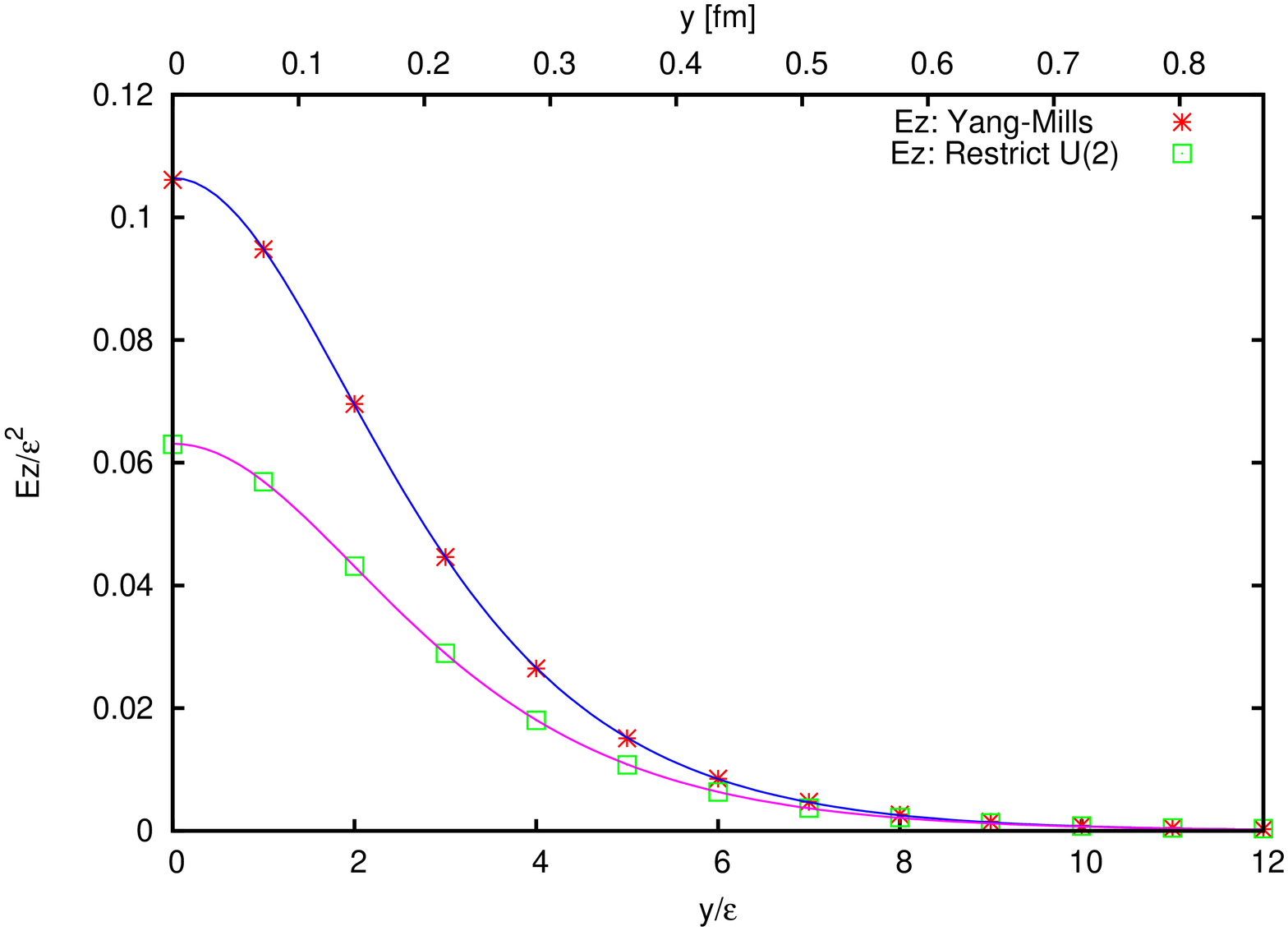} \includegraphics[,
width=5.0cm,
angle=270
]
{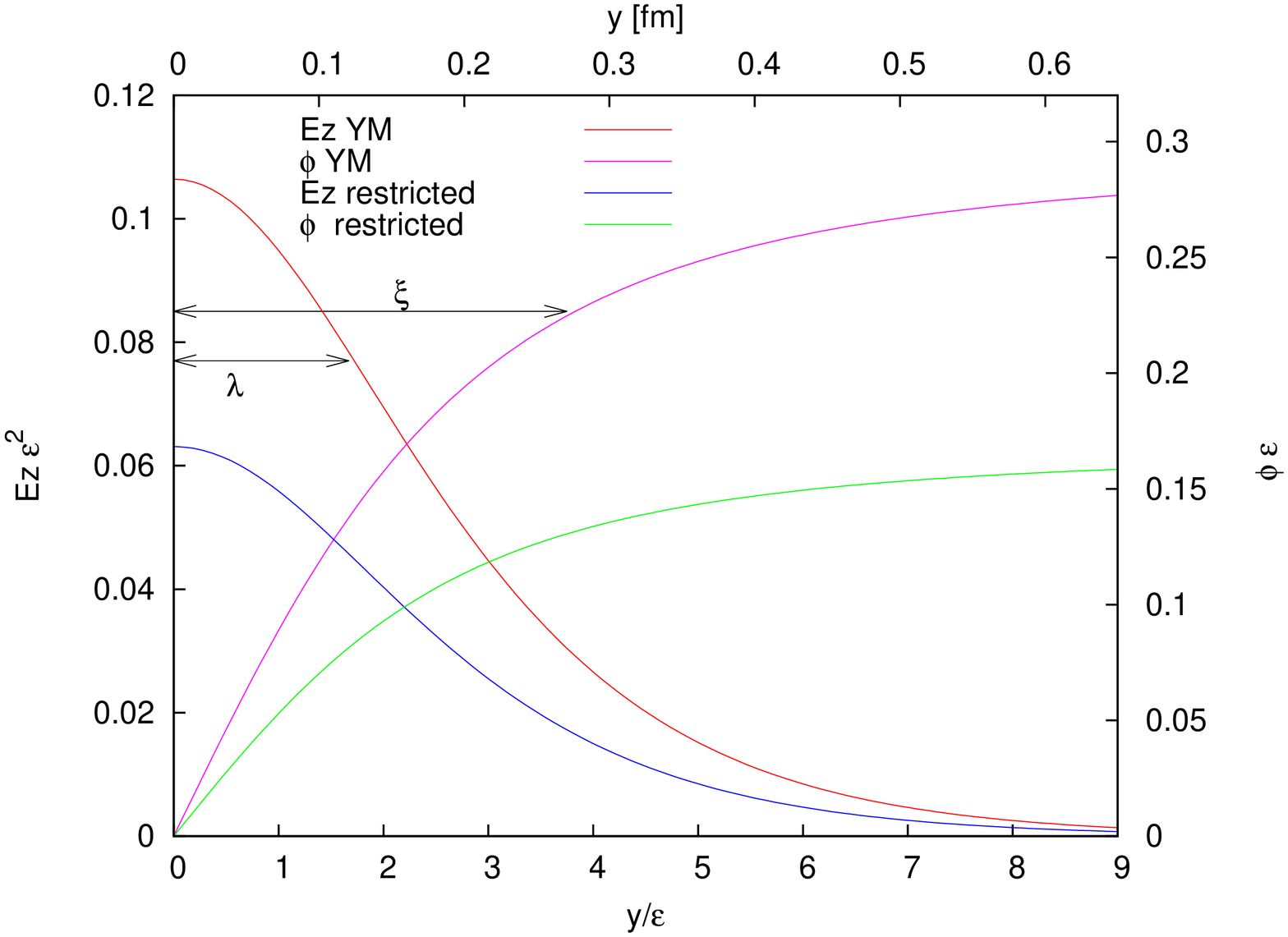}
\end{center}
\caption{ (Left panel) The plot of the chromoelectric field $E_{z}$ versus the
distance $y$ in units of the lattice spacing $\epsilon$ and the fitting as a
function $E_{z}(y)$ of $y$ according to ({\protect \ref{eq:fluxClem}}). The red cross
for the original $SU(3)$ field and the green square symbol for the restricted
field. (Right panel) The order parameter $\phi$ reproduced as a function
$\phi(y)$ of $y$ according to ({\protect \ref{order-f}}), togather with the
chromoelectric field {$E_{z}(y)$.}}%
\label{fig:type}%
\end{figure}

Moreover, we investigate the type of the QCD vacuum as the dual
superconductor. The left panel of Fig.\ref{fig:type} is the plot for the
chromoelectric field $E_{z}$ as a function of the distance $y$ in units of the
lattice spacing $\epsilon$ for the original $SU(3)$ field and for the
restricted field.

In order to examine the type of the dual superconductivity, we apply the
formula for the magnetic field derived by Clem \cite{Clem75} in the ordinary
superconductor based on the Ginzburg-Landau (GL) theory to the chromoelectric
field in the dual superconductor. In the GL theory, the gauge field $A$ and
the scalar field $\phi$ obey simultaneously the GL equation and the Ampere
equation:
\begin{subequations}
\begin{align}
(\partial^{\mu}-iqA^{\mu})(\partial_{\mu}-iqA_{\mu})\phi+\lambda(\phi^{\ast
}\phi-\eta^{2})  &  =0,\\
\partial^{\nu}F_{\mu\nu}+iq[\phi^{\ast}(\partial_{\mu}\phi-iqA_{\mu}%
\phi)-(\partial_{\mu}\phi-iqA_{\mu}\phi)^{\ast}\phi]  &  =0.
\end{align}

Usually, in the dual superconductor of the type II, it is justified to use the
asymptotic form $K_{0}(y/\lambda)$ to fit the chromoelectric field in the
large $y$ region (as the solution of the Ampere equation in the dual GL
theory). However, it is clear that this solution cannot be applied to the
small $y$ region, as is easily seen from the fact that $K_{0}(y/\lambda
)\rightarrow\infty$ as $y\rightarrow0$. In order to see the difference between
type I and type II, it is crucial to see the relatively small $y$ region.
Therefore, such a simple form cannot be used to detect the type I dual
superconductor. However, this important aspect was ignored in the preceding
studies except for a work \cite{Cea:2012qw}.

On the other hand, Clem \cite{Clem75} does not obtain the analytical solution
of the GL equation explicitly and use an approximated form for the scalar
field $\phi$ (given below in (\ref{order-f})). This form is used to solve the
Ampere equation exactly to obtain the analytical form for the gauge field
$A_{\mu}$ and the resulting magnetic field $B$. This method does not change
the behavior of the gauge field in the long distance, but it gives a finite
value for the gauge field even at the origin. Therefore, we can obtain the
formula which is valid for any distance (core radius) $y$ from the axis
connecting $q$ and $\bar{q}$: the profile of chromoelectric field in the dual
superconductor is obtained:
\end{subequations}
\begin{equation}
E_{z}(y)=\frac{\Phi}{2\pi}\frac{1}{\zeta\lambda}\frac{K_{0}(R/\lambda)}%
{K_{1}(\zeta/\lambda)},\text{ \qquad}R=\sqrt{y^{2}+\zeta^{2}},
\label{eq:fluxClem}%
\end{equation}
provided that the scalar field is given by (See the right panel of
Fig.\ref{fig:type})
\begin{equation}
\phi(y)=\frac{\Phi}{2\pi}\frac{1}{\sqrt{2}\lambda}\frac{y}{\sqrt{y^{2}%
+\zeta^{2}}}, \label{order-f}%
\end{equation}
where $K_{\nu}$ is the modified Bessel function of the $\nu$-th order,
$\lambda$ the parameter corresponding to the London penetration length,
$\zeta$ a variational parameter for the core radius, and $\Phi$ external
electric flux. In the dual superconductor, we define the GL parameter $\kappa$
as the ratio of the London penetration length $\lambda$ and the coherence
length $\xi$ which measures the coherence of the magnetic monopole condensate
(the dual version of the Cooper pair condensate): $\kappa=\lambda/\xi.$It is
given by \cite{Clem75} 
\begin{equation}
\kappa=\frac{\lambda}{\xi}=\sqrt{2}\frac{\lambda}{\zeta}\sqrt{1-K_{0}%
^{2}(\zeta/\lambda)/K_{1}^{2}(\zeta/\lambda)}.
\end{equation}
\begin{table}[th]%
\begin{tabular}
[c]{|l||c|c|c|c|c|}\hline
& $\lambda/\epsilon$ & $\zeta/\epsilon$ & $\xi/\epsilon$ & $\Phi$ & $\kappa
$\\\hline
SU(3) YM field & $1.672\pm0.014$ & $3.14\pm0.09$ & $3.75\pm0.12$ &
$4.36\pm0.3$ & $0.45\pm0.01$\\\hline
restricted field & $1.828\pm0.023$ & $3.26\pm0.13$ & $3.84\pm0.19$ &
$2.96\pm0.3$ & $0.48\pm0.02$\\\hline
\end{tabular}
\caption{ The properties of the Yang-Mills vacuum as the dual superconductor
obtained by fitting the data of chromoelectric field with the prediction of
the dual Ginzburg-Landau theory. }%
\label{Table:GL-fit}%
\end{table}

According to the formula Eq.(\ref{eq:fluxClem}), we estimate the GL parameter
$\kappa$ for the dual superconductor of $SU(3)$ YM theory, although this
formula is obtained for the ordinary superconductor of $U(1)$ gauge field.
Table~\ref{Table:GL-fit} shows the fitting result, and the left panel of
Figure \ref{fig:type} shows the obtained fitted functions for YM-field and the
restricted field. The superconductor is type I if $\kappa<\kappa_{c}$, while
type II if $\kappa>\kappa_{c}$, where the critical value of GL parameter
dividing the type of the superconductor is given by $\kappa_{c}=1/\sqrt
{2}\simeq0.707$. Our data clearly shows that the dual superconductor of
$SU(3)$ YM theory is type I with
\begin{equation}
\kappa_{YM}=0.45\pm0.01.
\end{equation}
This result is consistent with a quite recent result obtained independently by
Cea, Cosmai and Papa \cite{Cea:2012qw}. The London penetration length
$\lambda=0.1207(17)$ fm and the coherence length $\xi=0.2707(86)$fm is
obtained in units of the string tension $\sigma_{\text{phys}}=(440\text{MeV}%
)^{2}$, and data of lattice spacing is taken from the TABLE I in
Ref.\cite{Edward98}. Moreover, our result shows that the restricted part plays
the dominant role in determining the type of the non-Abelian dual
superconductivity of the $SU(3)$ YM theory, i.e., type I with
\begin{equation}
\kappa_{V}=0.48\pm0.02,
\end{equation}
$\lambda=0.132(3)$fm and $\xi=0.277(14)$fm. This is a novel feature overlooked
in the preceding studies. Thus the restricted-field dominance can be seen also
in the determination of the type of dual superconductivity where the
discrepancy is just the normalization of the chromoelectric field at the core
$y=0$, coming from the difference of the total flux $\Phi$. These are
gauge-invariant results. Note again that this restricted-field and the
non-Abelian magnetic monopole extracted from it reproduce the string tension
in the static quark--antiquark potential \cite{lattice2010,abeliandomSU(3)}.

Our result should be compared with the result obtained by using the Abelian
projection: Y. Matsubara et. al \cite{Matsubara:1993nq} suggests
$\kappa=0.5\sim1$ (which is $\beta$ dependent), border of type I and type II
for both $SU(2)$ and $SU(3)$. In $SU(2)$ case, on the other hand, there are
other works \cite{Suzuki:2009xy,Chernodub:2005gz} which conclude that the type
of vacuum is at the border of type I and type II.

\subsection{Confinement/deconfinement phase transition at  finite
temperature}

From now on, we investigate confinement/deconfinement phase transition at 
finite temperature in view of the dual super conductivity picture. We use
Polyakov loops as quark and antiquark source in place of the Wilson loop. By
measuring the chromo-flux created by the Polyakov loop pair for both the
YM-field and the restricted field (V-field):%
\begin{equation}
P_{U}(\vec{x}):=\mathrm{tr}\left(  \prod\nolimits_{t=1}^{N_{T}}U_{(\vec
{x},t),4}\right)  \text{, \ \ \ \ \ \ \ }P_{V}(x):=\mathrm{tr}\left(
\prod\nolimits_{t=1}^{N_{T}}V_{(\vec{x},t),4}\right)  ,
\end{equation}
we test the V-field dominance for the Polyakov loops for the various
temperature. Then, we investigate the chromo-flux and the phase transition in
view of the non-Abelian dual Meissner effect.

\subsubsection{Polyakov loops and their correlation functions}

\begin{figure}[ptb]
\begin{center}
\includegraphics[
height=5.5cm, angle=270]
{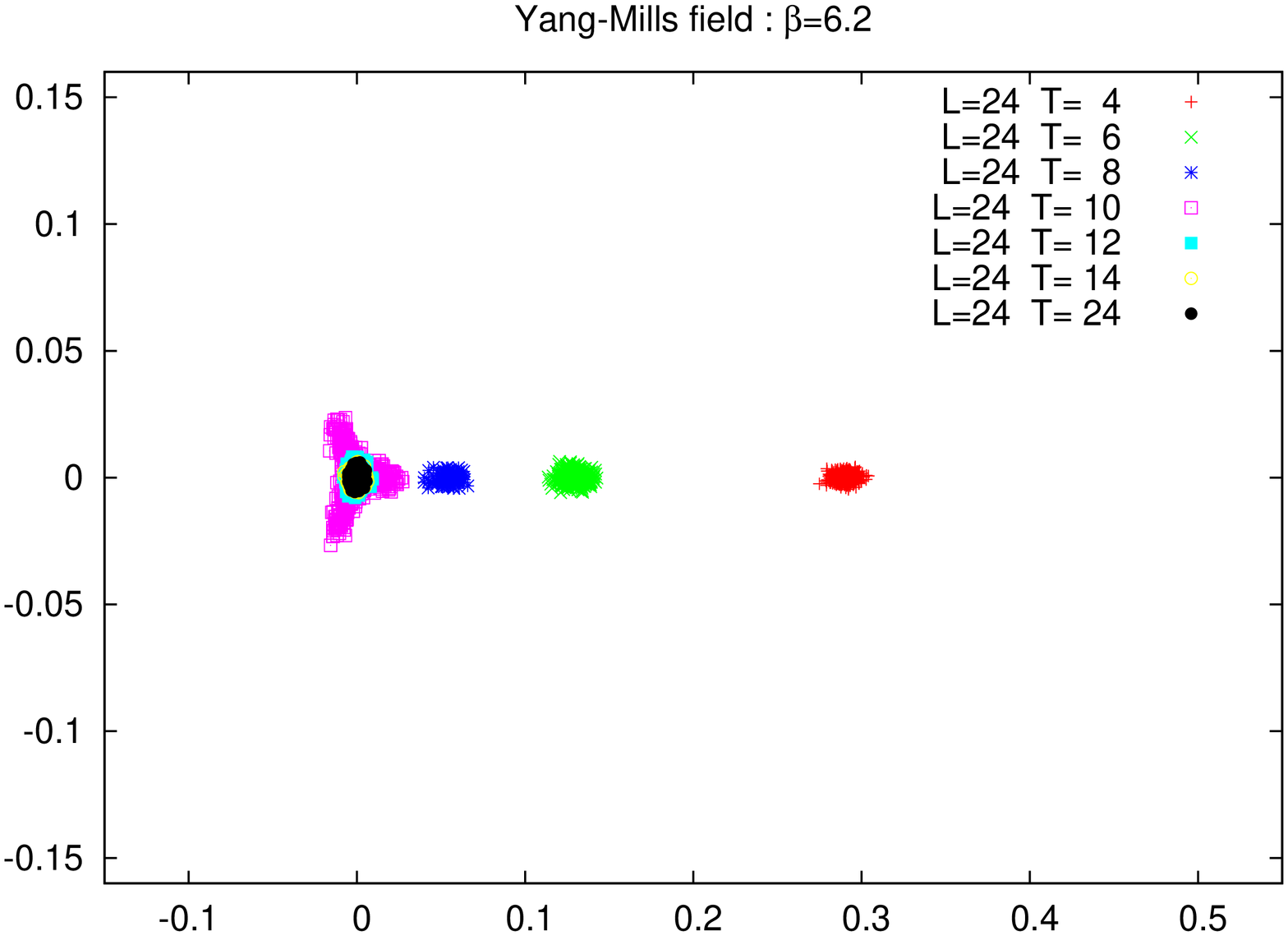} \ \ \includegraphics[
height=5.5cm, angle=270]
{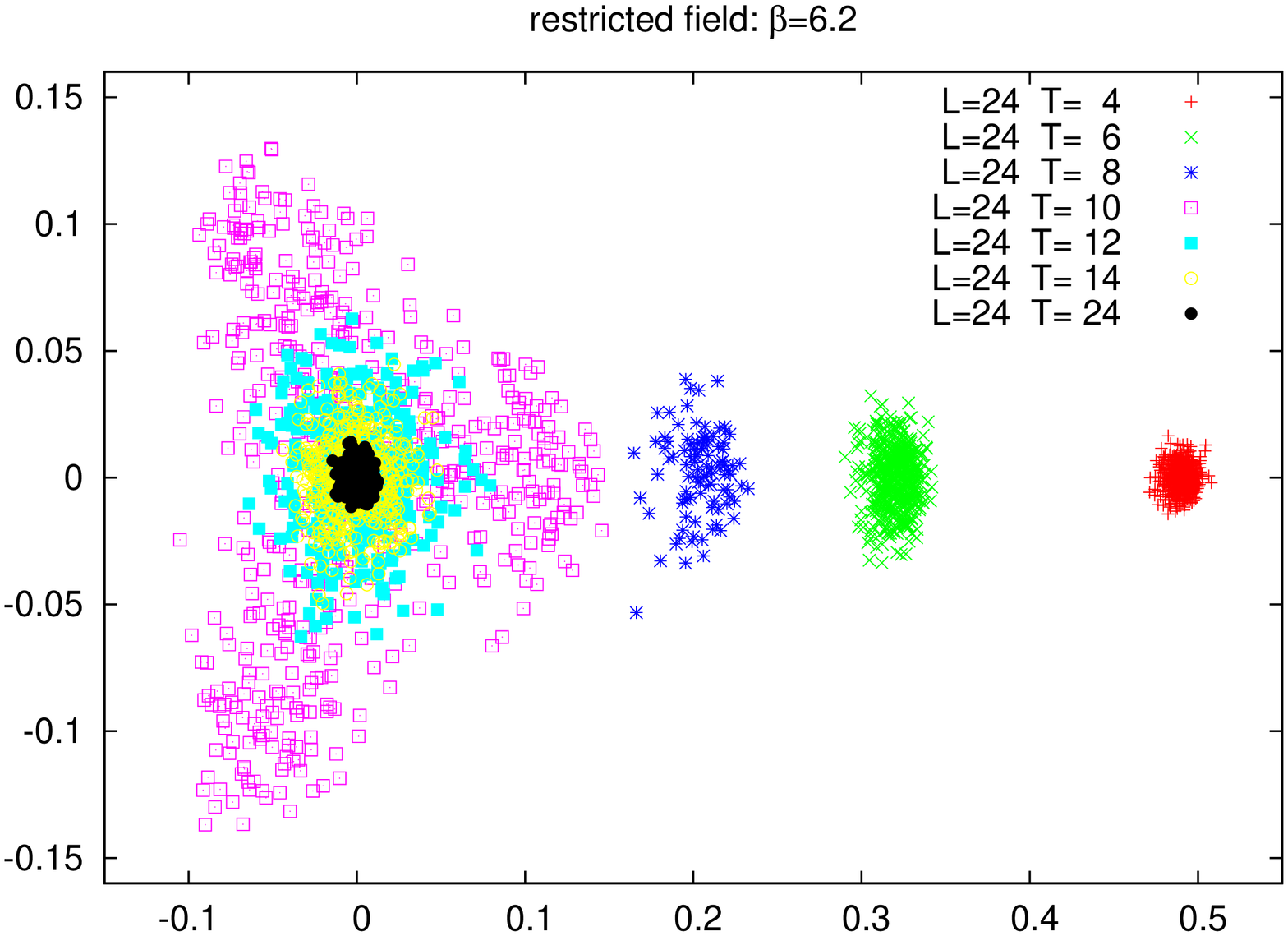}
\end{center}
\caption{{}The distribution of the space-averaged Polyakov loop for each
ocnfiguration, Eq.({\protect \ref{eq:PLP}})  (Left) For the YM field. (Right) For the
restricted field. }%
\label{Fig:PLP}%
\end{figure}

\begin{figure}[ptbh]
\begin{center}
\includegraphics[
height=5.5cm, angle=270]
{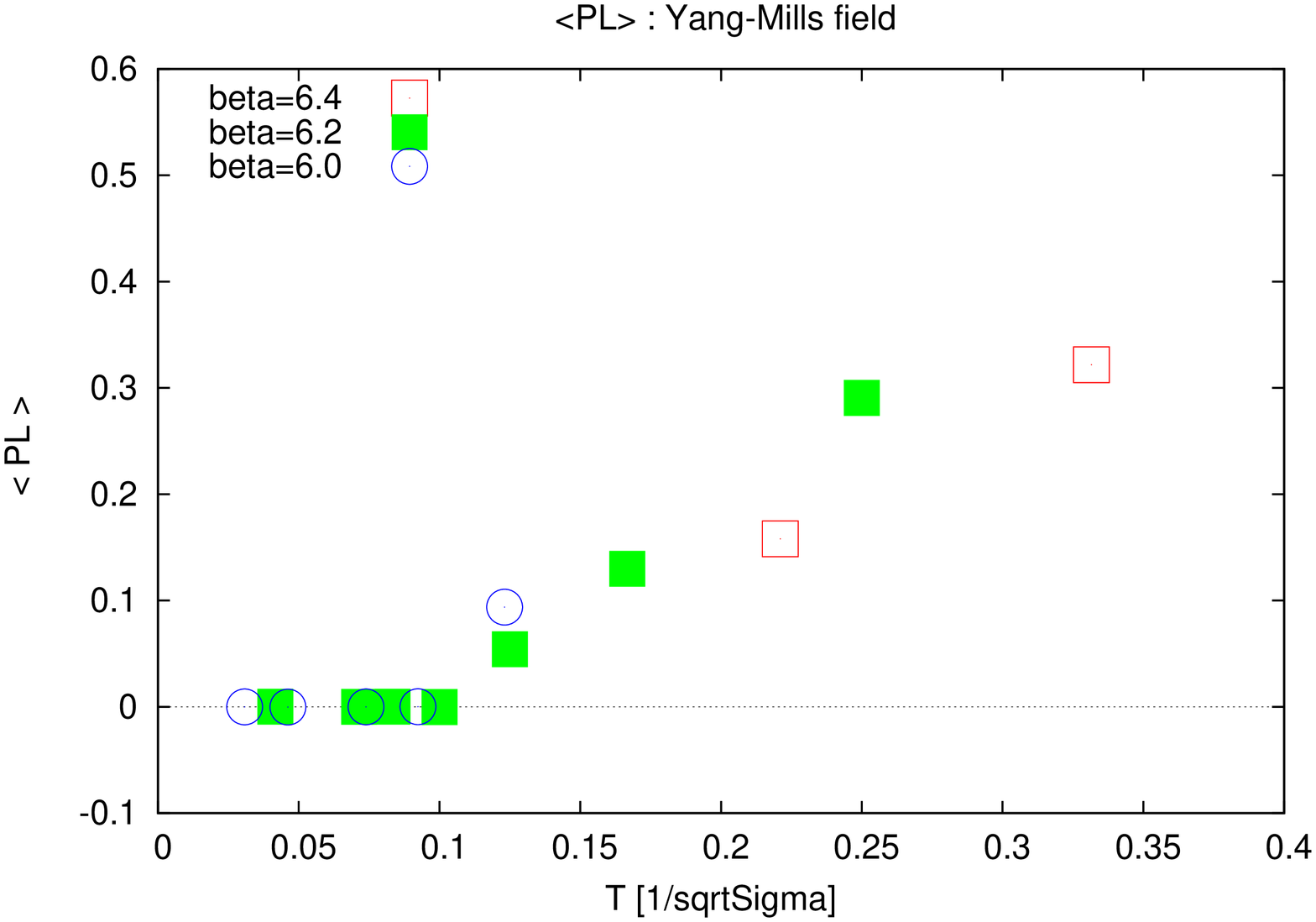} \ \ \includegraphics[
height=5.5cm, angle=270]
{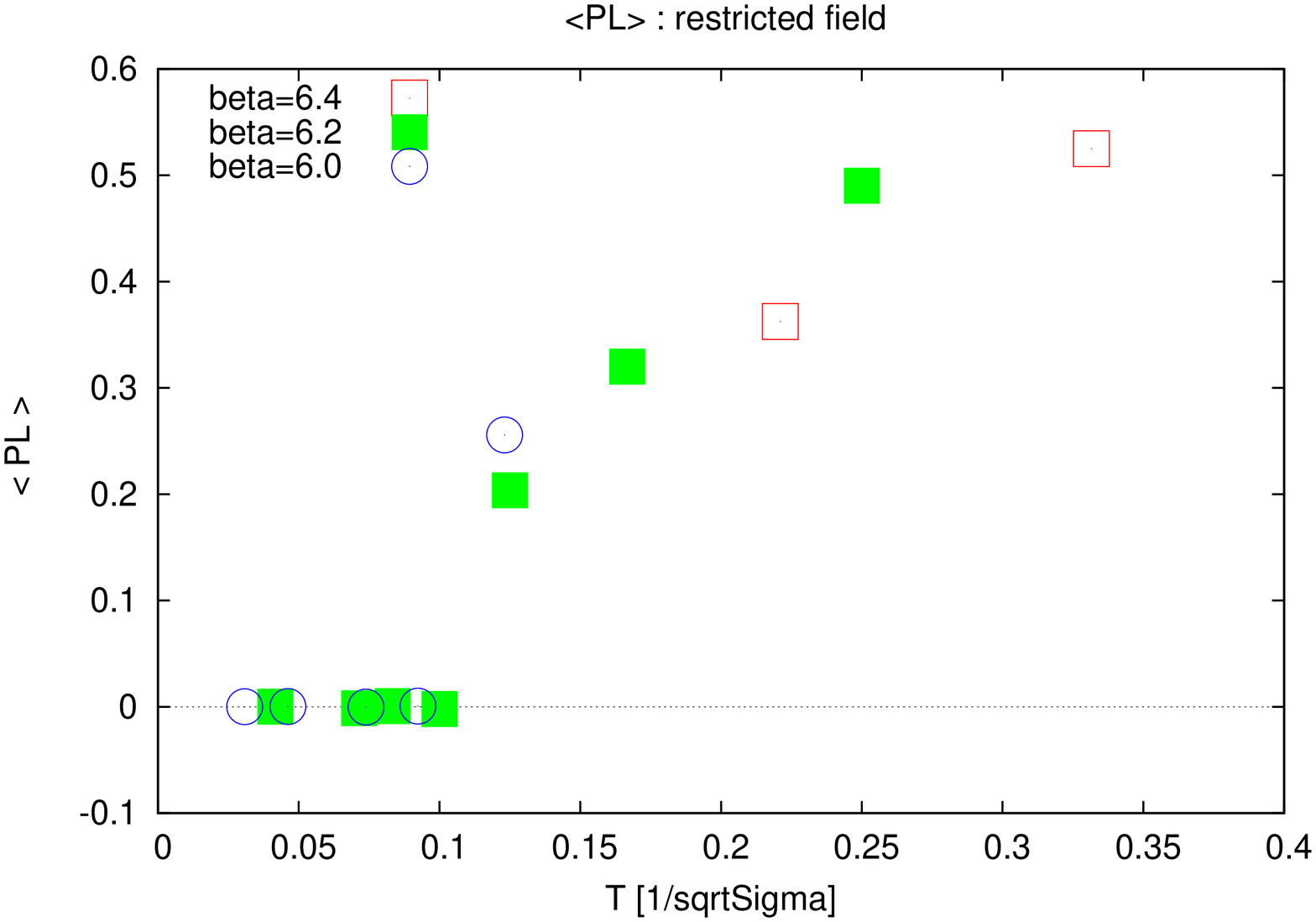}
\end{center}
\caption{{}The Ployakov loop average: (Left) For the YM field $\left\langle
P_{U}\right\rangle $ . (Right) For the restricted field $\left\langle
P_{V}\right\rangle .$}%
\label{fig:PLP-ave}%
\end{figure}

Figure \ref{Fig:PLP} show the distribution of space-averaged Polyakov loops
for each configuration:
\begin{equation}
P_{U}:=L^{-3}\sum_{\{\vec{x}\}}P_{U}(\vec{x}),\text{ \ \ \ }P_{V}:=L^{-3}%
\sum_{\{\vec{x}\}}P_{V}(x).\label{eq:PLP}%
\end{equation}
The left panel shows the distribution of YM\ field and the right panel that of
the V-field. We obtain the Polyakov loop average for configurations, which is
the conventional order parameter for confinement and deconfinement phase
transition in $SU(3)$ YM\ theory. Figure \ref{fig:PLP-ave} shows the Polyakov
loop average for the YM field $\left\langle P_{U}\right\rangle $ (left panel)
and restricted field $\left\langle P_{V}\right\rangle $ (right panel). Each
panel shows the same critical temperature of confinement/deconfinement phase
transition. 

\begin{figure}[ptb]
\begin{center}
\includegraphics[
height=5.3cm, angle=270]
{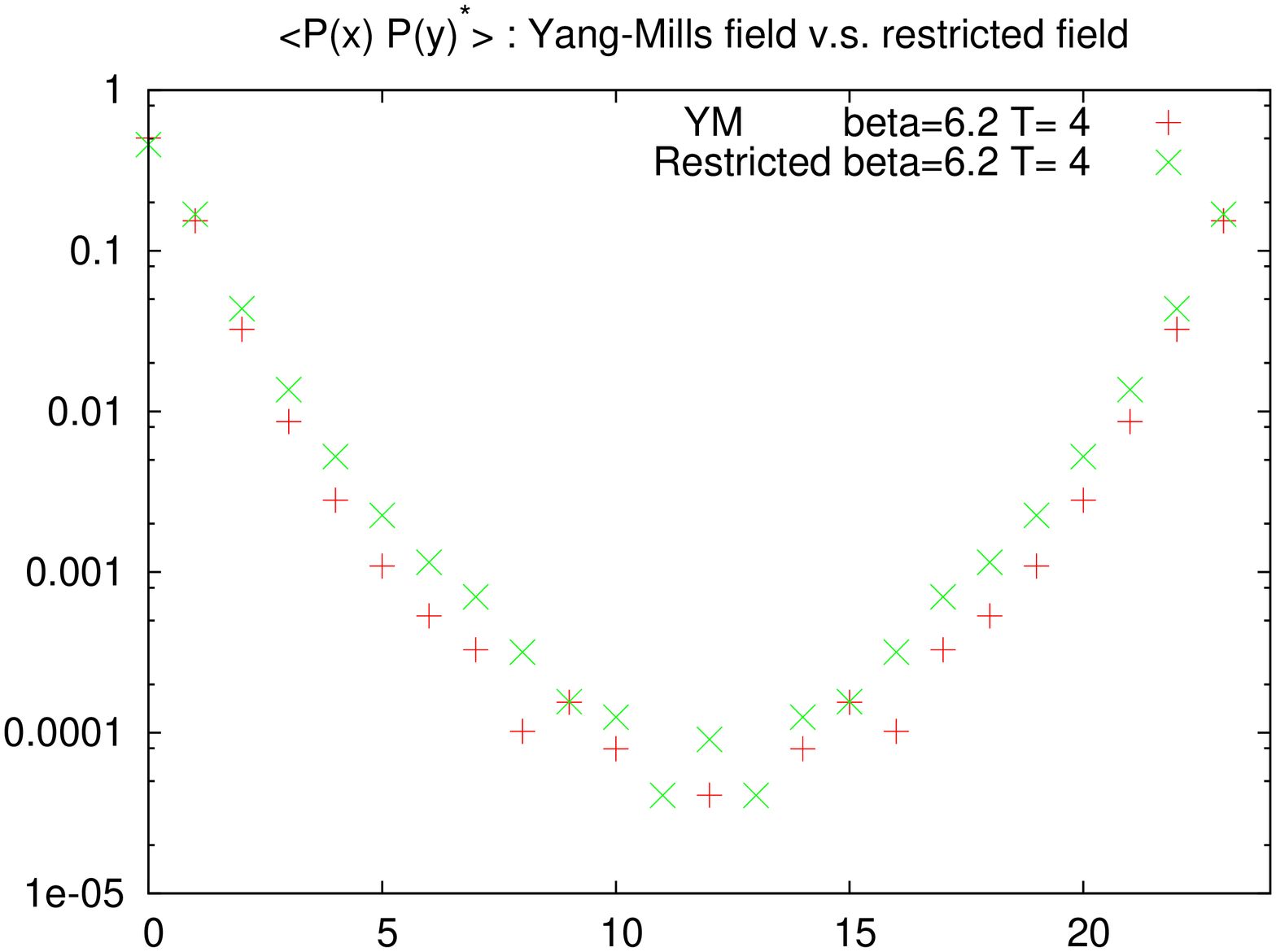}\includegraphics[
height=5.3cm, angle=270]
{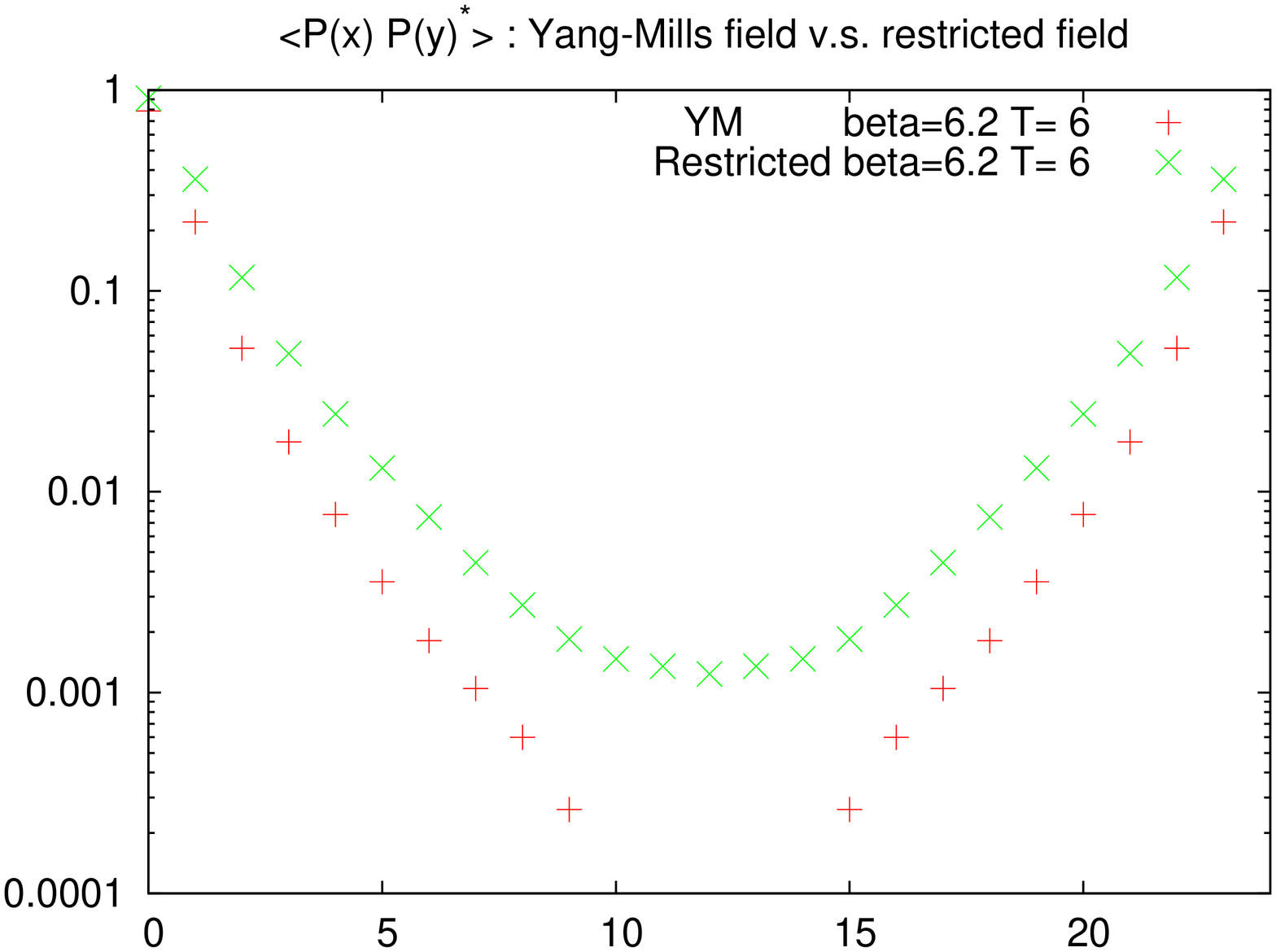}\includegraphics[
height=5.3cm, angle=270]
{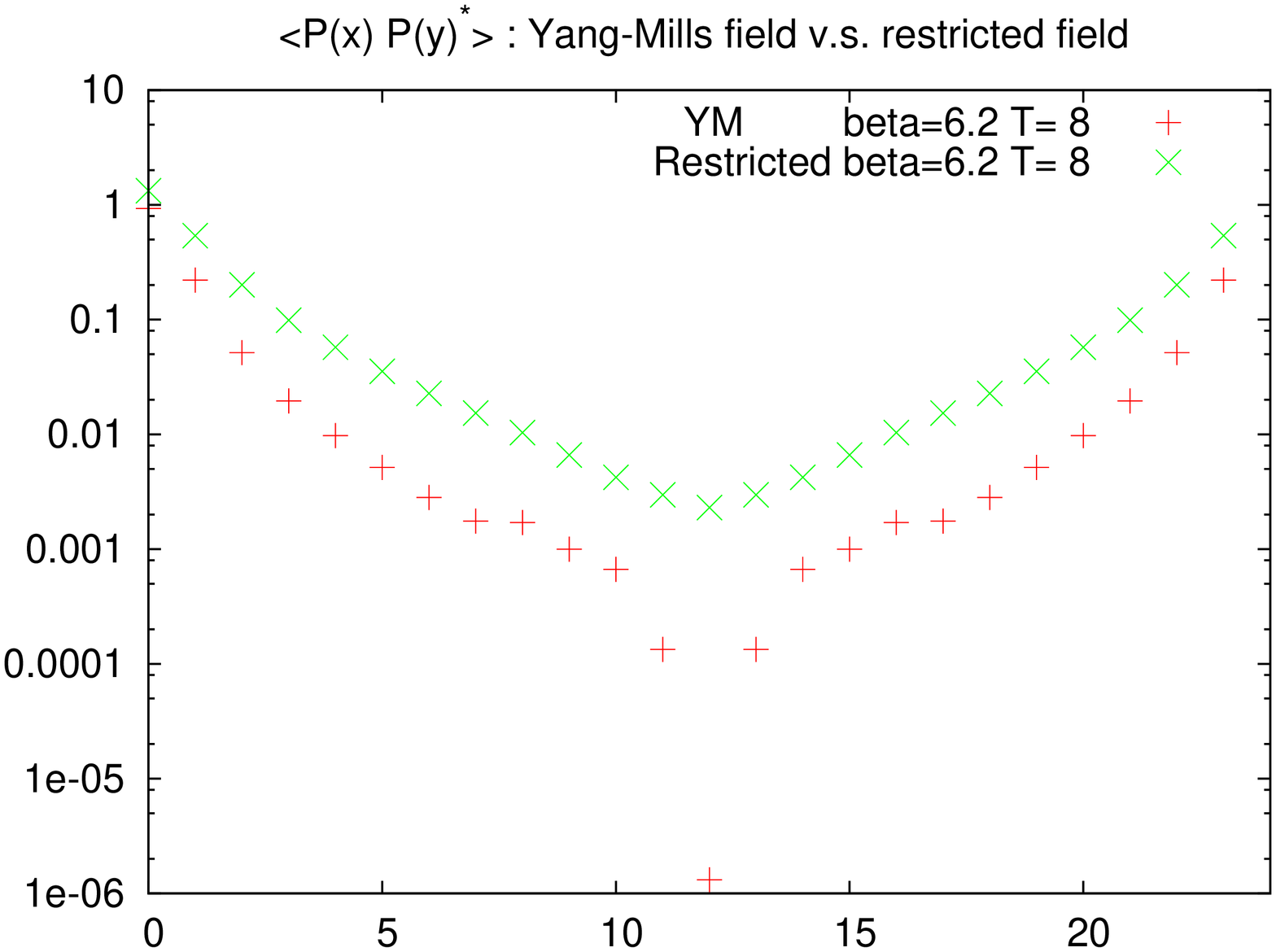} \includegraphics[
height=5.3cm, angle=270]
{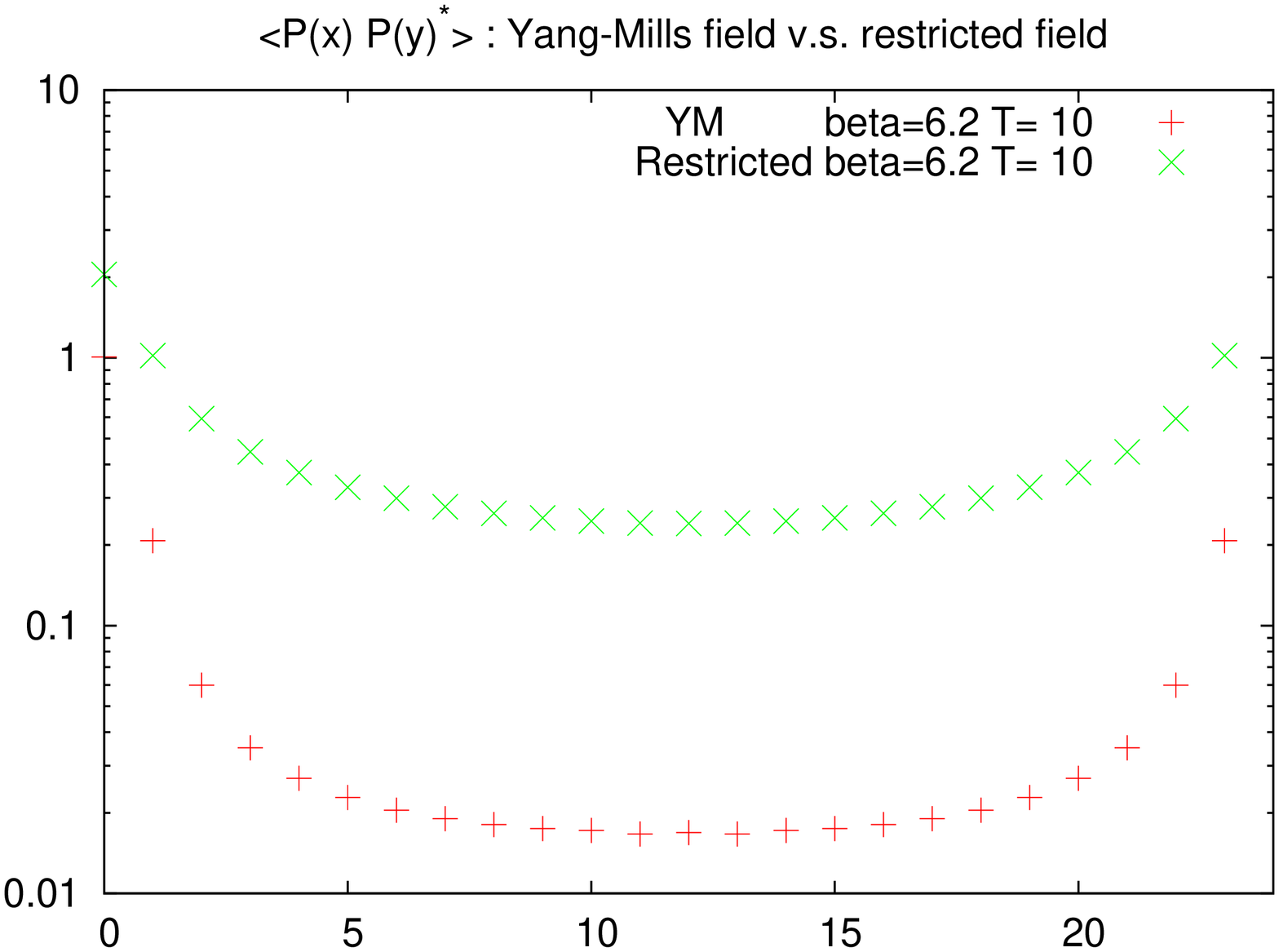} \includegraphics[
height=5.3cm, angle=270]
{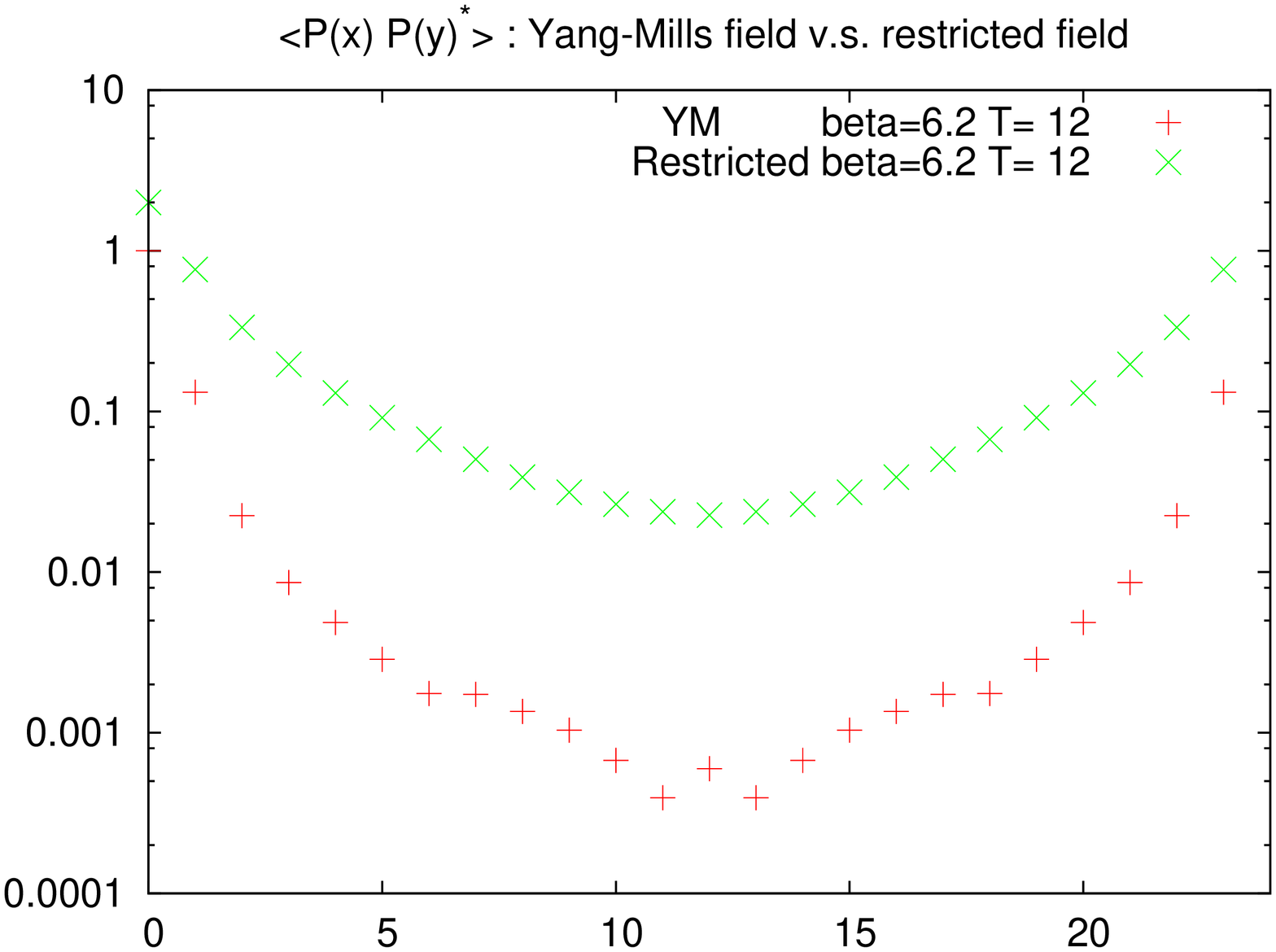} \ \ \includegraphics[
height=5.3cm, angle=270]
{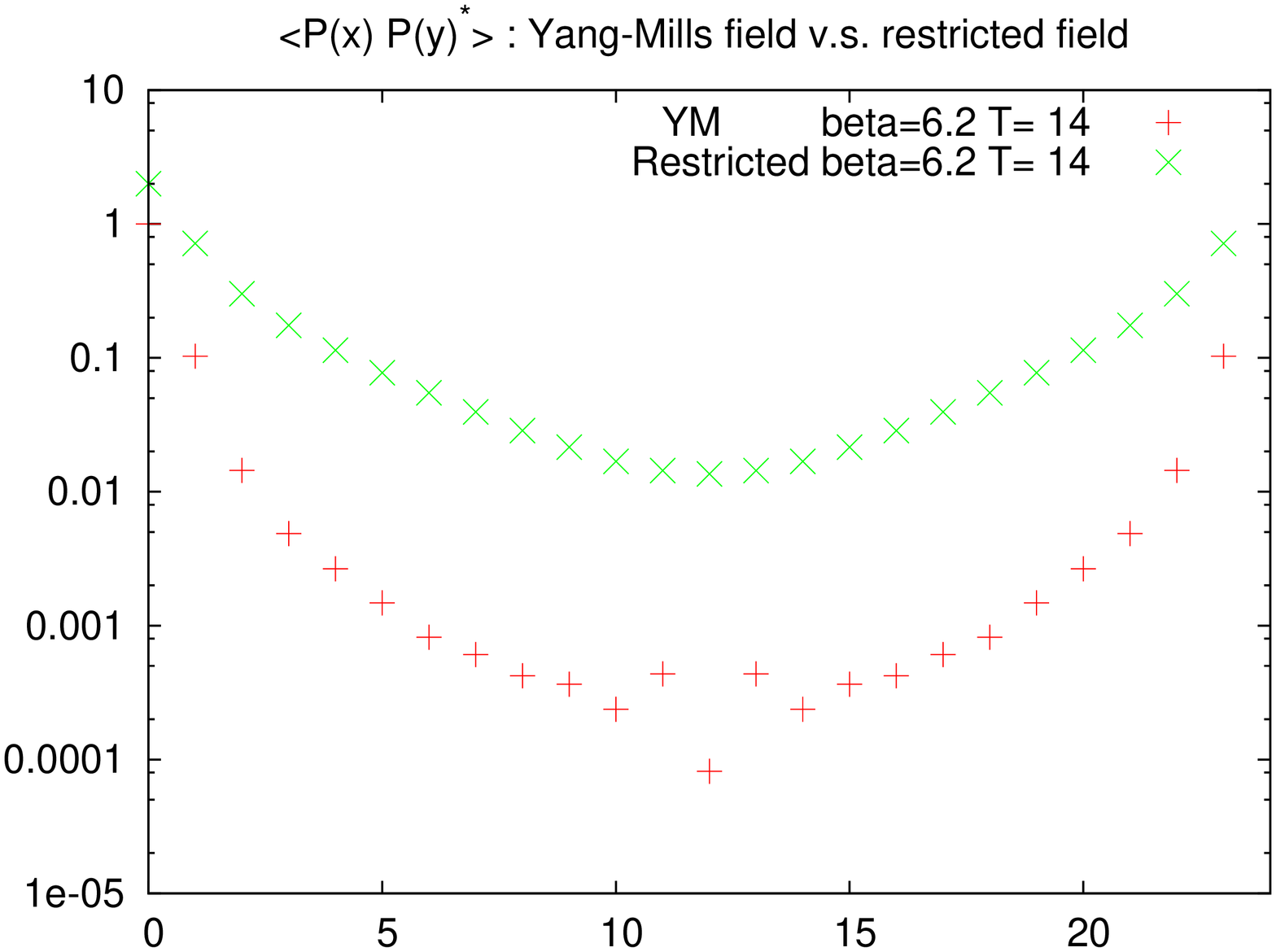}
\end{center}
\caption{Comparison of \ the correlation function of the Polyakov loop for the
YM field and the restriced field at various temperatures: The panels are
arranged from the upper-left to lower-right panel in order of the high to low
temperature.}%
\label{fig:PLP-correlations}%
\end{figure}

Then, we investigate two-point correlation function of Polyakov loop:%
\begin{equation}
D_{U}(x-y):=\left\langle P_{U}(x)^{\ast}P_{U}(y)\right\rangle -\left\langle
|P_{U}|^{2}\right\rangle ,\text{ \ \ \ }D_{V}(x-y):=\left\langle
P_{V}(x)^{\ast}P_{V}(y)\right\rangle -\left\langle |P_{V}|^{2}\right\rangle ,
\end{equation}
Figure \ref{fig:PLP-correlations} shows that the comparison of the
$D_{U}(x-y)$ and $D_{V}(x-y)$ for each temperature. Every panel shows that the
YM-field and restricted field ($V$-field) have the same profile, i.e., we can
extract the dominant mode for the quark confinement by the $V$%
-field.

\subsubsection{Chromoelectric flux in deconfinement phase}

Next, we investigate the non-Abelian dual Meissner effect at finite
temperature. To investigate the chromo flux, we use the gauge invariant
correlation function which is used at zero temperature. (see the left
panel\ of Fig. \ref{fig:Operator}). Note that at finite temperature, we must
use the operator with the same size in the temporal direction, and the quark
and antiquark pair is replaced by a pair of the Polyakov loop with the
opposite direction.

\begin{figure}[ptb]
\begin{center}
\includegraphics[
height=6cm, angle=270]
{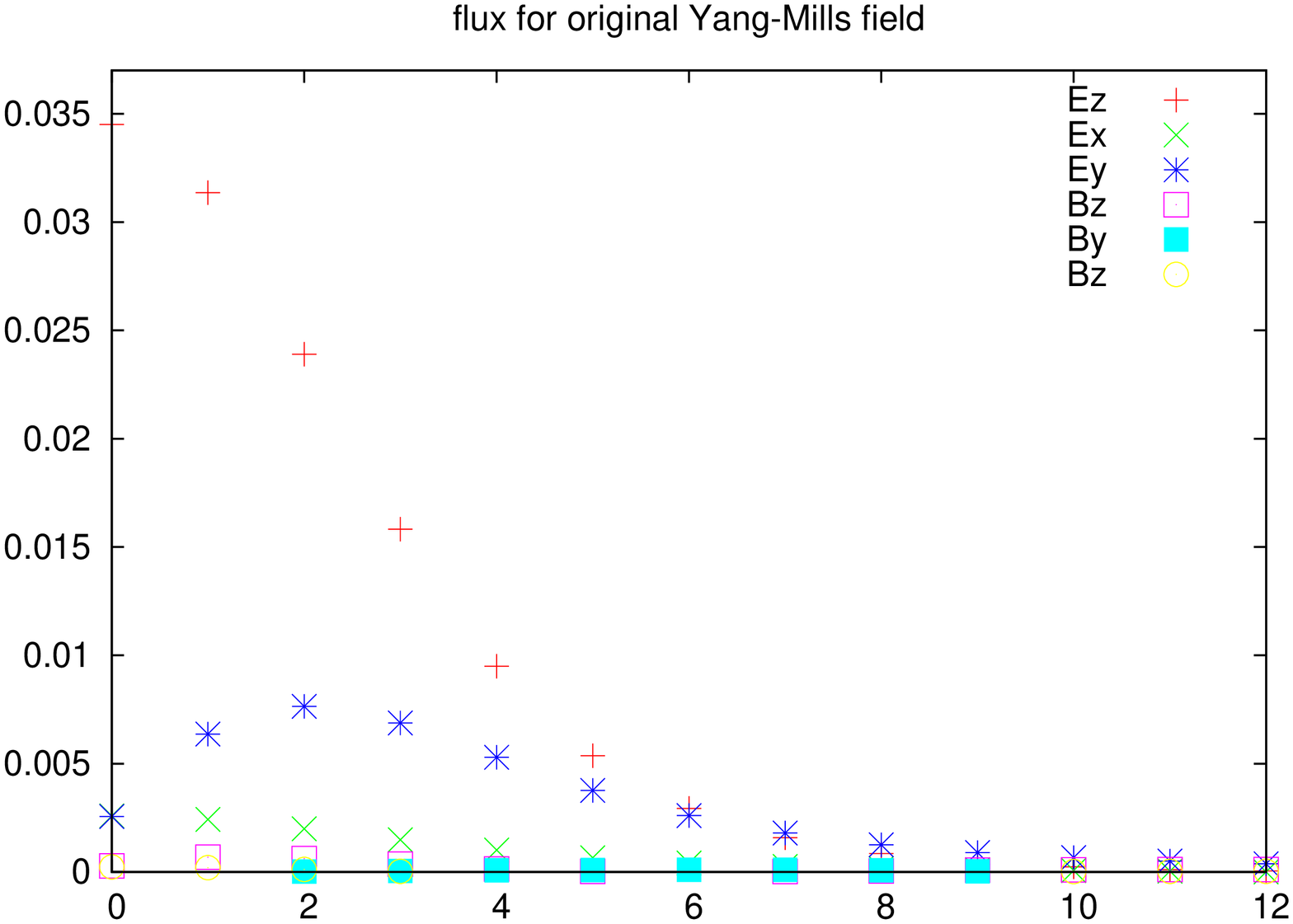} \ \ \ \ \ \ \includegraphics[
height=6cm, angle=270]
{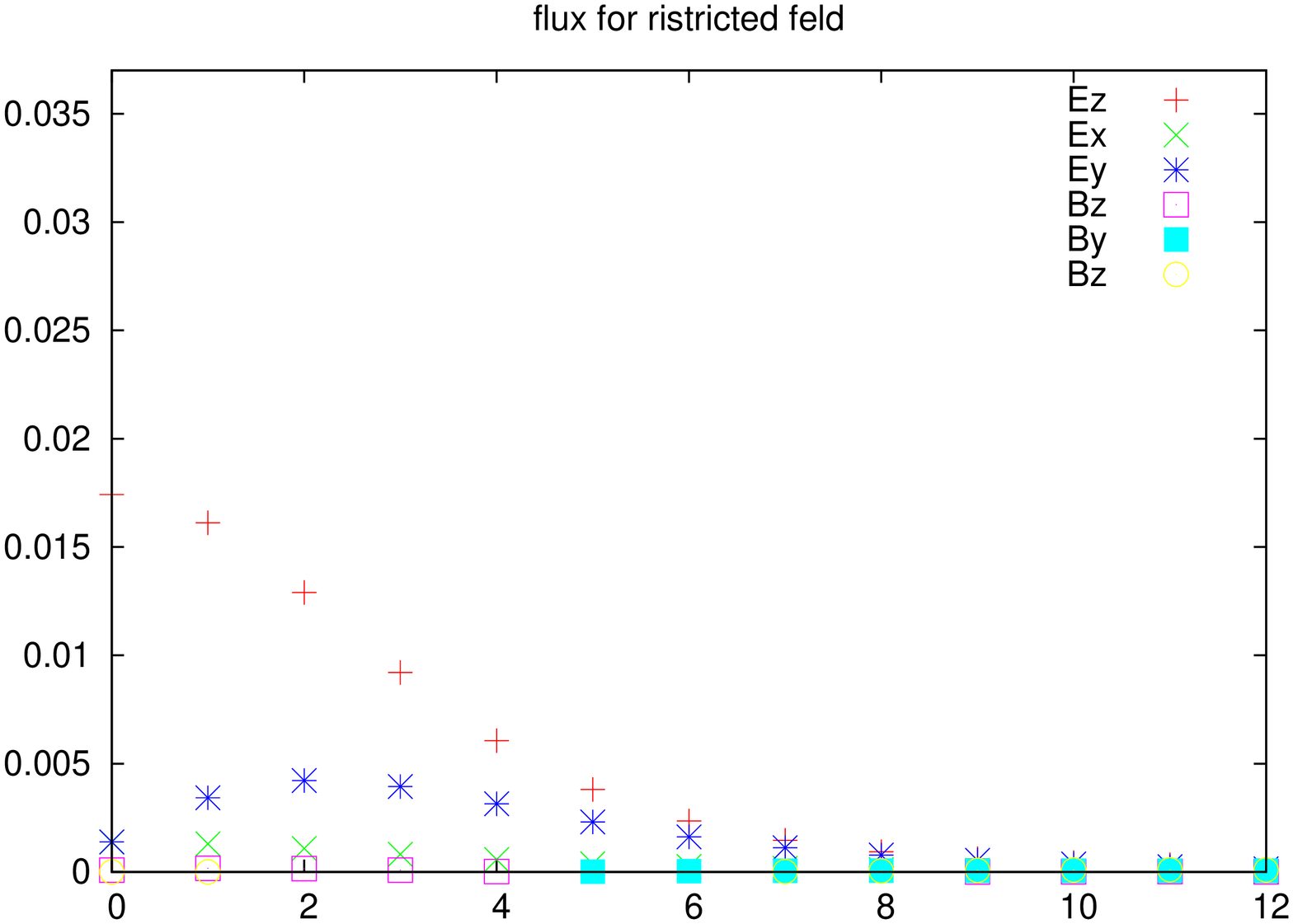}
\end{center}
\caption{{}The chromo-flux created by quark-antiquark pair in the plane
$z=1/3R$ for a quark at $z=0$ and an antiquark at $z=R$
(Fig.{\protect \ref{fig:Operator}}) by moving the probe, $U_{p}$ or $V_{p}$ along the
y-direction. (Left) For the YM-field (Right)\ For the restricted field.}%
\label{fig:flux-T}%
\end{figure}

Figure \ref{fig:flux-T} shows the measurement of chromoelectric and
chromomagnetic flux at high temperature $T>T_{c}$ $\ $(for the lattice
$N_{T}=6,$ $\beta=6.2$\.{)}. We measure the chromo-flux of quark-antiquark
pair in the plane $z=1/3R$ for a quark at $z=0$ and an antiquark at $z=R$
(Fig.\ref{fig:Operator}) by moving the probe, $U_{p}$ or $V_{p}$ along the
y-direction. We observe the chromoelectric flux tube only in the direction
connecting quark and antiquark pair, while the other components take vanishing
values. We can observe no more squeezing of the chromoelectric flux tube, but
non-vanishing $E_{y}$ component \ in the chromoelectric field, which must be
compared with the result at zero temperature: Fig.\ref{fig:measure}. This
shows the disappearance of the dual Meissner effect at high
temperature.

\section{Summary and outlook}

We have studied the dual superconductivity for $SU(3)$ YM theory by using our
new formulation of YM theory on a lattice. We have extracted the restricted
field ($V$-field) from the YM field which plays a dominant role in confinement
of quark (fermion in the fundamental representation) at finite temperature,
i.e., the restricted field dominance in Polyakov loop. Then we have measured
the chromoelectric and chromomagnetic flux for both the original YM field and
the restricted field at low temperature in confinement phase. We have observed
evidences of the dual Messier effect of $SU(3)$ YM theory, i.e., the
chromoelectric flux tube and the associated non-Abelian magnetic monopoles
created by quark and antiquark pair.

At high temperature ($T>T_{c}$) in the deconfinement phase, we have observed
the disappearance of the dual Meissner effect by measuring the chromo flux.
\ Note that, the Polyakov loop average cannot be the direct signal of the dual
Meissner effect or magnetic monopole condensation. Therefore, it is important
to find a order parameter which detects the dual Meissner effect directly, and
to investigate whether the order parameter in view of the dual Meissner effect
gives the same critical temperature with that of the Polyakov loop average. At
low temperature in confinement phase, we have observed non-vanishing
magnetic(-monopole) current created by the quark-antiquark source,
$\mathbf{k}$, Eq(\ref{def-k}), therefore, $\mathbf{k}$ can be the order
parameter of confinement/deconfinement transition in view of dual Meissner
effect. We are now under investigation on it and the result will appear in the
near future.

\subsection*{Acknowledgement}

This work is supported by Grant-in-Aid for Scientific Research (C) 24540252
from Japan Society for the Promotion Science (JSPS), and also in part by JSPS
Grant-in-Aid for Scientific Research (S) 22224003. The numerical calculations
are supported by the Large Scale Simulation Program No.12-13 (FY2012), No.
12/13-20 (FY2012/13) of High Energy Accelerator Research Organization (KEK).


\begin{thebibliography}{99}                                                                                               %


\bibitem {dualSC}{\small Y. Nambu, Phys. Rev. D10, 4262 (1974); G. 't Hooft,
in: High Energy Physics, edited by A.; Zichichi (Editorice Compositori,
Bologna, 1975); S. Mandelstam, Phys. Report 23, 245 (1976); A.M. Polyakov,
Nucl. Phys. B120, 429 (1977).}

\bibitem {tHooft81}{\small G. 't Hooft, Nucl. Phys. B190, 455 (1981).}

\bibitem {Suzuki90}{\small T. Suzuki and I. Yotsuyanagi, Phys. Rev. D42, 4275
(1990).}

\bibitem {stack94Shiba}{\small J.D. Stack, S.D. Neiman and R.Wensley, Phys.
Rev. D50, 3399 (1994); H. Shiba and T. Suzuki, Phys. Lett. B351 519 (1995).}

\bibitem {greensite}{\small J. Greensite, Prog. Part. Nucl. Phys. 51 1
(2003).}

\bibitem {SCGTKKS08L}{\small K.-I. Kondo, A.Shibata, T. Shinohara, T.
Murakami, S. Kato and S. Ito, Phys. Lett. B669, 107 (2008).}

\bibitem {exactdecomp}{\small A. Shibata, K.-I. Kondo and T. Shinohara,
arXiv:0911.5294[hep-lat], Phys. Lett. B691, 91-98 (2010).}

\bibitem {CFNS-C}{\small Y.M. Cho, Phys. Rev. D 21, 1080 (1980). Phys. Rev. D
23, 2415 (1981); Y.S. Duan and M.L. Ge, Sinica Sci. 11, 1072 (1979); L.
Faddeev and A.J. Niemi, Phys. Rev. Lett. 82, 1624 (1999); S.V. Shabanov, Phys.
Lett. B 458, 322 (1999). Phys. Lett. B 463, 263 (1999).}

\bibitem {KSM05}{\small K.-I. Kondo, T. Murakami and T. Shinohara, Eur. Phys.
J. C 42, 475 (2005); K.-I. Kondo, T. Murakami and T. Shinohara, Prog. Theor.
Phys. 115, 201 (2006).}

\bibitem {Cho80c}{\small Y.M. Cho, Phys. Rev. Lett. 44, 1115(1980).}

\bibitem {FN99a}{\small L. Faddeev and A.J. Niemi, Phys. Lett. B 449, 214
(1999). \ Phys. Lett. B 464, 90(1999).}

\bibitem {SCGTKKS08}{\small K.-I. Kondo, T. Shinohara and T. Murakami, Prog.
Theor. Phys. 120, 1 (2008).}

\bibitem {KKMSSI06}{\small S. Kato, K.-I. Kondo, T. Murakami, A. Shibata, T.
Shinohara, and S. Ito, hep-lat/0509069, Phys. Lett. B632, 326-332 (2006).}

\bibitem {ref:NLCVsu2}{\small S. Ito, S. Kato, K.-I. Kondo, A. Shibata, and T.
Shinohara, Phys. Lett. B645, 67--74 (2007).}

\bibitem {ref:NLCVsu2-2}{\small A. Shibata, S. Kato, K.-I. Kondo, T. Murakami,
T. Shinohara, S. Ito, Phys.Lett. B653 101-108 (2007).}

\bibitem {lattce2007}{\small A. Shibata, S. Kato, K.-I. Kondo, T. Shinohara
\ and S. Ito, POS(LATTICE2007) 331, arXiv:0710.3221 [hep-lat]}

\bibitem {lattice2008}{\small A. Shibata, S. Kato, K.-I. Kondo, T. Shinohara
and S. Ito, 56 [hep-lat], PoS(LATTICE 2008) 268}

\bibitem {kato:lattice2009}{\small S. Kato, K-I. Kondo, A. Shibata and T.
Shinohara, PoS(LAT2009) 228.}

\bibitem {lattice2009}{\small A. Shibata, K.-I. Kondo, S. Kato, S. Ito, T.
Shinohara, N. Fukui, PoS LAT2009 (2009) 232, arXiv:0911.4533[hep-lat].}

\bibitem {lattice2010}{\small A. Shibata, K.-I. Kondo, S. Kato and T.
Shinohara, PoS(Lattice 2010)286}

\bibitem {lattice2012}{\small A. Shibata, K.-I. Kondo, S. Kato and T.
Shinohara, PoS LATTICE2012 (2012) 215.}

\bibitem {confinmentX}{\small A. Shibata, K.-I. Kondo, S. Kato and T.
Shinohara, PoS ConfinementX (2012) 052.}

\bibitem {DMeisner-TypeI2013}{\small A. Shibata, K-I. Kondo, S. Kato and T.
Shinohara, Phys.Rev. D87 (2013) 5, 054011, arXiv:1212.6512 }

\bibitem {kondo:taira:2000}{\small K.-I. Kondo and Y. Taira, e-Print:
hep-th/9911242, Prog. Theor. Phys. 104, 1189-1265 (2000). e-Print:
hep-th/9906129, Mod. Phys. Lett .15, 367-377 (2000). Nucl. Phys. Proc.Suppl.
83, 497-499 (2000).}

\bibitem {KondoNAST}{\small K.-I. Kondo, Phys. Rev. D77, 085029 (2008).}

\bibitem {KondoShibata}{\small K.-I. Kondo and A. Shibata,
arXiv:0801.4203[hep-th], CHIBA-EP-170, KEK-PREPRINT-2007-73}

\bibitem {Giacomo}{\small A. Di Giacomo, M. Maggiore, and S. Olejnik, Phys.
Lett. B236, 199 (1990); Nucl. Phys. B347, 441 (1990).}

\bibitem {flusx:AP}{\small Y. Matsubara, S. Ejiri and T. Suzuki, NPB Poc.
suppl~34, 176~(1994)}

\bibitem {Suganuma}{\small S. Gongyo, T. Iritani and H. Suganuma, Phys. Rev.
D86, 094018 (2012). H. Suganuma, K. Amemiya, H. Ichie, N. Ishii, H. Matsufuru,
T. Takahashi, hep-lat/0407016, Nucl. Phys. B (Proc. Suppl.) 106, 679-681
(2002).}

\bibitem {abeliandomSU(3)}{\small K.-I. Kondo, A. Shibata, T. Shinohara, and
S. Kato, Phys. Rev. D83, 114016 (2011).}

\bibitem {APEsmear}{\small M. Albanese et al. (APE Collaboration), Phys. Lett.
B 192,163 (1987).}

\bibitem {Edward98}{\small R.G. Edwards, U.M. Heller and T.R. Klassen, Phys.
Rev. Lett. 80, 3448--3451 (1998).}

\bibitem {Clem75}{\small J.R. Clem, J. Low. Temp. Phys. 18, 427 (1975).}

\bibitem {Matsubara:1993nq}{\small Y.~Matsubara, S.~Ejiri and T.~Suzuki,
Nucl.\ Phys.\ Proc.\ Suppl.\ 34, 176 (1994) [hep-lat/9311061].}

\bibitem {Cea:2012qw}{\small P.~Cea, L.~Cosmai and A.~Papa, Phys.\ Rev.\ D 86,
054501 (2012) [arXiv:1208.1362 [hep-lat]].}

\bibitem {Suzuki:2004dw}{\small T.~Suzuki, K.~Ishiguro, Y.~Mori and T.~Sekido,
Phys.\ Rev.\ Lett.\ 94, 132001 (2005) [hep-lat/0410001]. }

\bibitem {Chernodub:2005gz}{\small M.~N.~Chernodub, K.~Ishiguro, Y.~Mori,
Y.~Nakamura, M.~I.~Polikarpov, T.~Sekido, T.~Suzuki and V.~I.~Zakharov,
Phys.\ Rev.\ D72, 074505 (2005), hep-lat/0508004. }

\bibitem {Suzuki:2009xy}{\small T.~Suzuki, M.~Hasegawa, K.~Ishiguro, Y.~Koma
and T.~Sekido, Phys.Rev.\ D80, 054504(2009) arXiv:0907.0583 [hep-lat];
K.~Ishiguro, M.~Hasegawa, Y.~Koma, T.~Sekido and T.~Suzuki, PoS LAT 2009, 238
(2009). }
\end{thebibliography}
\end{document}